\documentclass[superscriptaddress,reprint,PRApplied,twocolumn]{revtex4-2}
\usepackage{graphicx}
\usepackage{dcolumn}
\usepackage{bm}
\usepackage{amsmath}
\usepackage{amssymb}
\usepackage{bm}
\usepackage{epsfig}
\usepackage{color}
\usepackage{subfigure}
\usepackage{wrapfig}
\usepackage{float}
\usepackage{hyperref}
\usepackage{enumitem}
\usepackage{natbib}
\usepackage{color}
\usepackage{siunitx}
\usepackage{makecell}
\hypersetup{
colorlinks,%
citecolor=blue,%
filecolor=blue,%
linkcolor=blue,%
urlcolor=black }

\makeatletter
\def\blfootnote{\gdef\@thefnmark{}\@footnotetext}
\makeatother

\begin{document}
\title{Charge stability and charge-state-based spin readout of shallow nitrogen-vacancy centers in diamond}
\author{Rakshyakar Giri $^\ddagger$}
\email{ragiri@dtu.dk}
\author{Rasmus H. Jensen $^\ddagger$}
\blfootnote{$^\ddagger$ These authors contributed equally to this work}
\author{Deepak Khurana}
\author{Juanita Bocquel}
\author{Ilya P. Radko}
\affiliation{Center for Macroscopic Quantum States (bigQ), Department of Physics, Technical University of Denmark, 2800 Kgs. Lyngby, Denmark}%
\author{Johannes Lang}
\author{Christian Osterkamp}
\author{Fedor Jelezko}
\affiliation{Institute for Quantum Optics and Center for Integrated Quantum Science and Technology (IQST), Ulm University, Albert-Einstein-Allee 11, 89081, Ulm, Germany}%
\author{Kirstine Berg-S{\o}rensen}
\affiliation{Department of Health Technology, Technical University of Denmark, 2800 Kgs. Lyngby, Denmark}%
\author{Ulrik L. Andersen}
\author{Alexander Huck}
\email{alhu@dtu.dk}
\affiliation{Center for Macroscopic Quantum States (bigQ), Department of Physics, Technical University of Denmark, 2800 Kgs. Lyngby, Denmark}%

\date{\today}

\begin{abstract}
Spin-based applications of the negatively charged nitrogen-vacancy (NV) center in diamonds require efficient spin readout. One approach is the spin-to-charge conversion (SCC), relying on mapping the spin states onto the neutral (NV$^0$) and negative (NV$^-$) charge states followed by a subsequent charge readout. With high charge-state stability, SCC enables extended measurement times, increasing precision and minimizing noise in the readout compared to the commonly used fluorescence detection. Nano-scale sensing applications, however, require shallow NV centers within a few $\si{\nano \meter}$ distance from the surface where surface related effects might degrade the NV charge state. In this article, we investigate the charge state initialization and stability of single NV centers implanted $\approx \SI{5}{\nano \meter}$ below the surface of a flat diamond plate. We demonstrate the SCC protocol on four shallow NV centers suitable for nano-scale sensing, obtaining a reduced readout noise of 5--6 times the spin-projection noise limit. We investigate the general applicability of SCC for shallow NV centers and observe a correlation between NV charge-state stability and readout noise. Coating the diamond with glycerol improves both charge initialization and stability. Our results reveal the influence of the surface-related charge environment on the NV charge properties and motivate further investigations to functionalize the diamond surface with glycerol or other materials for charge-state stabilization and efficient spin-state readout of shallow NV centers suitable for nano-scale sensing.
\end{abstract}


\maketitle

\section{Introduction}
The negatively charged nitrogen-vacancy (NV$^-$) center in diamond has excellent spin properties at room temperature with a long lifetime and coherence, enabling a broad range of applications. These include the sensing of magnetic and electric fields with high sensitivity and nanometer-scale spatial resolution~\cite{Taylor2008, Balasubramanian2008, Dolde2011}. Central to any NV$^-$ center sensing task is the ability to initialize and read out the spin state optically. The most common approach to read out the spin state is via the fluorescence rate, enabled by spin-dependent intersystem crossing from the triplet excited to a singlet state~\cite{Doherty2013}. The intersystem crossing mechanism sets a lower limit on the spin readout duration to around $\SI{300}{\nano\second}$ as determined by the singlet-state lifetime~\cite{Steiner2010} that, together with low photon counts in the order of $0.1$ per readout hinders the efficient measurement of the spin state. In the context of sensing, it is thus required to repeat the scheme multiple times in order to build up sufficient photon statistics to determine the spin state~\cite{Steiner2010, Jiang2009}. Nanopillars~\cite{Hedrich2020} or solid immersion lenses~\cite{Wildanger2012} might be used to increase the photon collection efficiency and improve the readout signal-to-noise ratio (SNR), but yet the attainable photon count rates are not sufficient to measure the spin state in a single shot.

In addition to photoelectrical readout~\cite{Siyushev2019}, another approach for spin readout involves the charge degree of freedom of the NV center, that in addition to NV$^{-}$ may also exist in the neutral charge state NV$^{0}$. Deep in the bulk, the charge state ratio between the two is highly stable in the dark or under weak optical excitation~\cite{Dhomkar2018, Bluvstein2019, Gorrini2021}.  With zero-phonon lines at $\SI{575}{\nano\meter}$ and $\SI{637}{\nano\meter}$ for NV$^{0}$ and NV$^{-}$, respectively, and therefore different absorption spectra, the charge states are distinguishable via selective illumination that excites NV$^{-}$ but not NV$^{0}$~\cite{Aslam2013}. In the spin-to-charge conversion (SCC) protocol~\cite{Shields2015}, population of the m$_{s}$ = 0 spin state is transferred to NV$^{0}$ via a photoionization process~\cite{Manson2005, Aslam2013}, while the population of the m$_{s}$ = $\pm1$ spin state remains in the NV$^-$ charge state. The subsequent charge-state readout may be long and efficient provided sufficient photon counts and charge-state stability is attained within the readout duration. Using this approach, $>1$ photon per readout may be detected when the system is in m$_{s}$ = $\pm1$, significantly increasing SNR per readout at the cost of significantly increased integration time as compared to standard fluorescence detection.  

Applying the SCC protocol to an NV center in a diamond nanobeam with significantly improved photon collection efficiency compared to bulk diamond, Shields et al. \cite{Shields2015} demonstrated an SNR $\approx 0.55$ per readout corresponding to a noise factor of $\approx 3 \sigma_p$, where $\sigma_p$ is the spin-projection noise limit~\cite{Hopper2018a}. More recently, the SCC readout has been demonstrated for NV center ensembles in nano-~\cite{Hopper2018} and bulk diamonds~\cite{Jayakumar2018}, and for single NV centers deep in the bulk~\cite{Jaskula2019}, and applied in covariance magnetometry~\cite{Rovny2022}.
Nano-scale sensing tasks including nuclear magnetic resonance (NMR)~\cite{Mamin2013,Staudacher2013} or sensing of free radicals~\cite{Martinez2020} however require single NV centers within a few $\SI{}{\nano \meter}$ from the diamond surface where the sensing signal is sufficiently strong~\cite{Muller2014}. In the vicinity of the surface, effects including upward energy band bending~\cite{Hauf2011} and electron tunneling to nearby surface traps~\cite{Bluvstein2019} are known to cause ionization of NV$^-$ to NV$^0$ at rates in the order of the inverse readout time or even faster. Although it has been demonstrated that SCC can work for NV centers implanted within a few $\SI{}{\nano\meter}$ from the surface in diamond nano-pillars~\cite{Ariyaratne2018}, the general applicability of the SCC protocol to near-surface NV centers as well as the possible detrimental impact of surface-related effects remains unexplored as of yet. 

In this article, we explore the charge-state stability and the SCC protocol on single NV centers implanted $\approx \SI{5}{\nano\meter}$ below the surface of a flat diamond plate. The implanted side of the diamond is accessible to apply a solution bath or nano-objects, an experimental configuration relevant for nano-scale sensing and spin mechanical coupling. We use a simplified experimental approach with only two lasers, one at $\SI{532}{\nano\meter}$ for NV charge and spin initialization and another at $\SI{594}{\nano\meter}$ for spin-state selective ionization and charge-state readout. From the detailed characterization of 33 single NV centers, we observe a strong variability of the probability to initialize into NV$^{-}$. We present data of four NV centres labelled NV$_{\text{A}}$ -- NV$_{\text{D}}$ that are representative of the distribution.
We show that after coating the diamond with deuterated glycerol, the initialization probability into NV$^-$ increases while simultaneously the ionization rate from NV$^-$ to NV$^0$ decreases as compared to the pristine surface. Furthermore, we demonstrate the general applicability of the SCC protocol on four near-surface NV centers. Our investigations indicate the importance of the local charge environment for the NV charge-state stability as well as the overall performance of the SCC protocol. We obtain a noise level as low as $\approx 5 \sigma_p$ corresponding to a single-shot SNR $\approx 0.3$, a result  comparable to that obtained with NV centers in the bulk or nano-structures with higher attainable photon count rates. 

\section{Experimental methods}
\begin{figure}[!htbp]
\centering
\includegraphics[width=1\columnwidth]{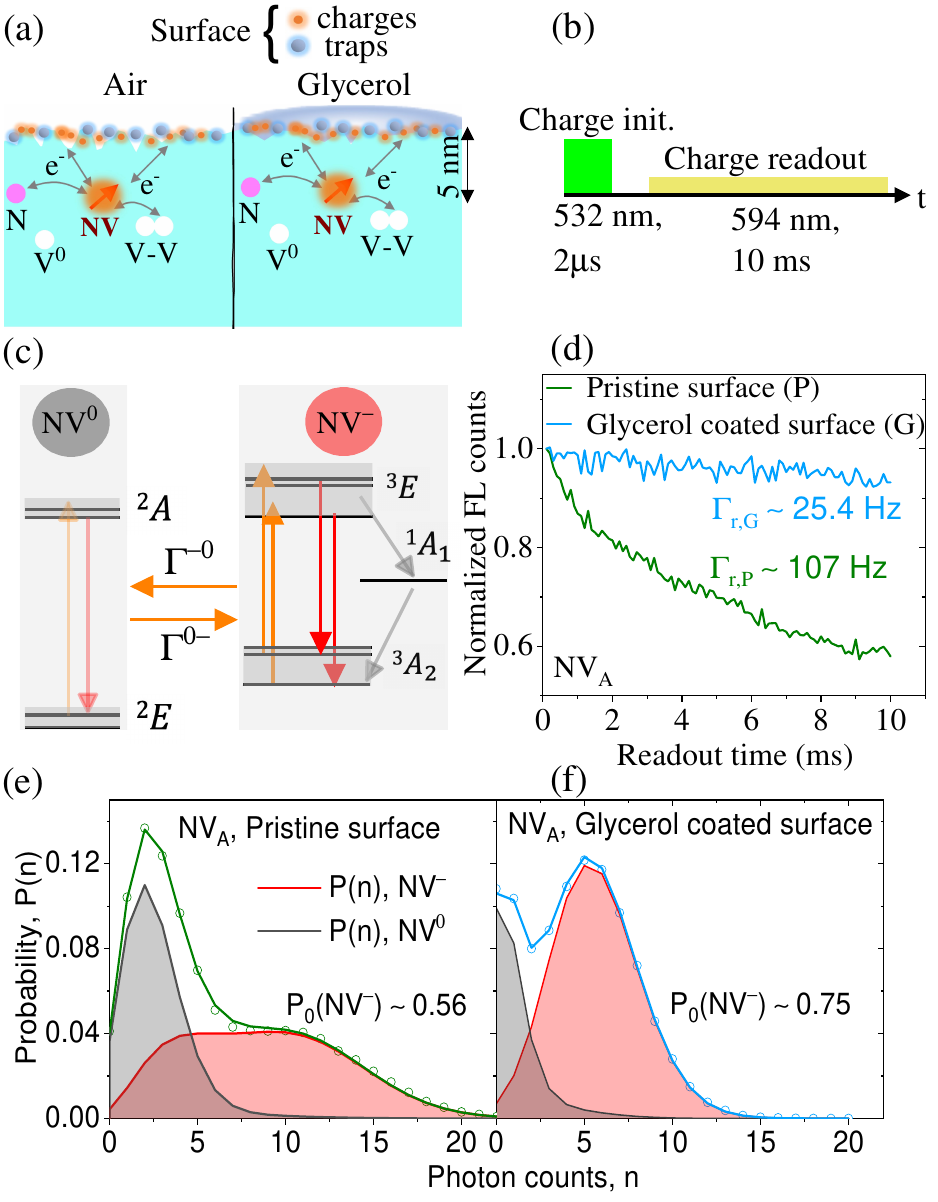}
\caption{Surface-related effects and charge-state stability of shallow NV centers. (a) Illustration of $\approx \SI{5}{\nano \meter}$ shallow NV centers and their interaction with the local charge environment with glycerol passivating surface charges and traps.  (b) Pulse sequence to measure the NV$^-$ charge-initialization probability P$_0$(NV$^{-}$), applying $\SI{532}{\nano\meter}$ initialization and $\SI{594}{\nano\meter}$ readout pulses, respectively. (c) Illustration of the energy levels of NV$^0$ and NV$^-$ together with allowed transition rates. The $\SI{594}{\nano \meter}$ laser ideally only excites NV$^{-}$. For strong excitation, it can ionize ($\Gamma^{-0}$) NV$^{-}$ to NV$^{0}$ and induce recombination ($\Gamma^{0-}$) from NV$^{0}$ to NV$^{-}$. The charge-state switching is suppressed for weak excitation, and the NV$^{0}$ center remains dark under 594~nm excitation. (d) Fluorescence evolution of NV$_{\text{A}}$ during the readout pulse, showing a reduced relaxation rate after adding glycerol ($\Gamma_{r,G}$) as compared to the relaxation with the pristine surface ($\Gamma_{r,P}$). (e, f) Photon-number distribution $P(n)$ of NV$_{\text{A}}$ with the pristine (e) and the glycerol-coated (f) surface. The gray- and red-shaded areas illustrate the respective contributions of NV$^{0}$ and NV$^{-}$.}
\label{Fig1}
\end{figure}

All measurements reported in this article were conducted on a 50-$\SI{}{\micro\meter}$-thin electronic-grade diamond plate (Element Six) with (100) surface orientation. The sample was overgrown with a 100-$\SI{}{\nano \meter}$-thick and $^{12}$C-enriched ($<$ 0.05$\%$ $^{13}$C) layer of diamond using chemical vapor deposition. NV centers were created via implantation of $^{15}$N$^{+}$ ions with an energy of 2.5~keV, and a dose of $5\times10^{8} \SI{}{\centi \meter}^{-2}$ (process pressure  $<$ $1\times10^{-7}$mbar)~\cite{Lang2020}. Subsequent annealing in ultra-high vacuum at $\SI{1000}\degreeCelsius$ for 3 hours heals radiation damage due to the ion bombardment, mobilizes vacancies, and through capture of an implanted nitrogen, yields a Gaussian-like distribution of single NV centers with a mean depth $\approx \SI{5}{\nano\meter}$ below the surface. The diamond was cleaned in a boiling acid mixture consisting in equal parts of nitric, perchloric, and sulfuric acid (tri-acid cleaning) to remove surface contaminants~\cite{Brown2019} and creating a hydrophilic surface with partial oxygen termination~\cite{Salvadori2010, Cui2013}. With these near-surface NV centers and using a Hahn-echo sequence, we measured T$_{2}$ coherence times in the range from $\numrange{3}{15} ~\SI{}{\micro \second}$, in good agreement with values obtained for a different sample fabricated using the same equipment and process parameters~\cite{Findler2020}. 

We used a home-built scanning confocal microscope setup with an NA=1.4 immersion-oil objective to investigate the NV charge state and implement the SCC protocol.
The power of each laser, $\SI{532}{\nano \meter}$ and $\SI{594}{\nano \meter}$, was controlled using acousto-optic modulators in a double-pass configuration to generate pulses for charge and spin initialization, shelving, ionization and readout. 
After long-pass filtering with a cut-on at $\SI{650}{\nano\meter}$, the NV fluorescence signal was detected with a single-photon-sensitive avalanche photodiode with dark counts $<\!\!200$ per second. Together with the selective excitation at $\SI{594}{\nano \meter}$, this configuration ensures, to a very high degree, that the signal stems from the NV$^-$ emission. With a motorized 3-axis translation stage we controlled the position of a permanent magnet to produce a bias magnetic field of $\approx \SI{7}{\milli T}$ along one of the four [111] crystal axes that all NV centers investigated here are aligned along. 
The diamond plate was horizontally mounted on a 170-$\SI{}{\micro\meter}$-thick cover glass with the NV layer on the top side, exposed either to air or solution and suitable for nanoscale sensing~[Fig.~\ref{Fig1}(a)]. The excitation lasers pass from below through the cover glass and the diamond, and the fluorescence is collected from the same side (see Supporting Information). 
Measurements on all the NV centers with a pristine surface and with a glycerol coated surface were performed without removing the diamond sample from the setup. To passivate the diamond surface, we carefully put a small droplet of deuterated glycerol (Glycerol-d$_8$ from Sigma-Aldrich) with a micro-pipette at the amount sufficient to cover the investigated region of diamond. No special procedure has been used to ensure spreading of glycerol through the surface, so the thickness of the layer was given by the glycerol surface tension, and was of the order of $\SI{100}{\micro\meter}$. 

\section{Results}
\subsection{Charge-state initialization and stability}
\begin{figure}[htbp!]
\centering
\includegraphics [width=1\columnwidth]{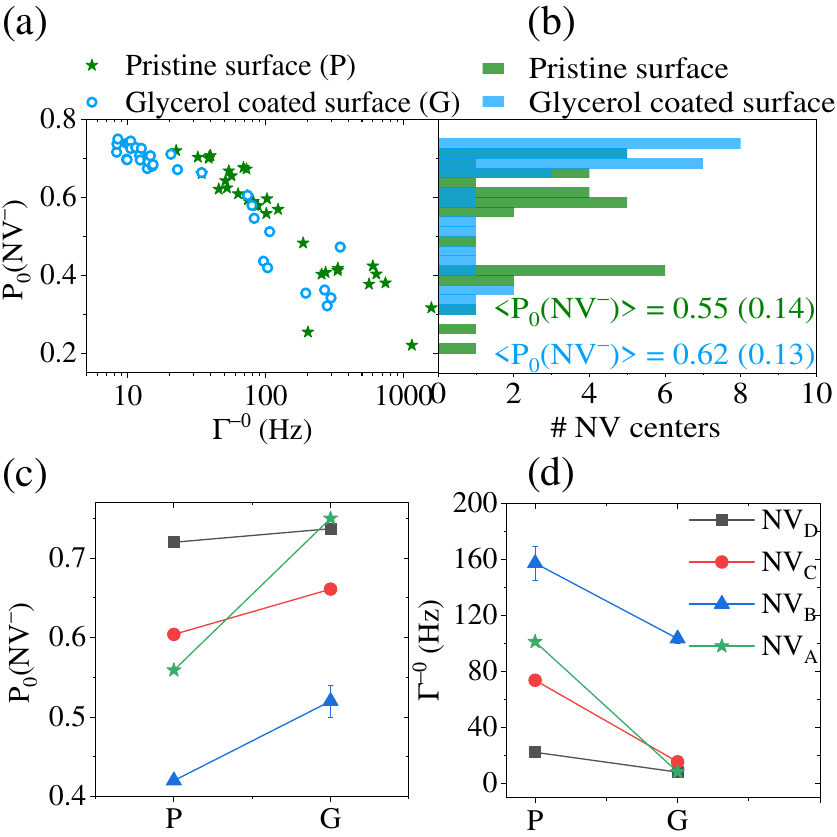}
\caption{(a) NV$^{-}$ charge initialization probability P$_0$(NV$^{-}$) versus ionization rate $\Gamma^{-0}$ for 33 NV centers with the pristine and the glycerol-coated surfaces. (b) Histogram of P$_0$(NV$^{-}$) (c, d) Comparison of P$_0$(NV$^{-}$) and $\Gamma^{-0}$ for pristine (P) and glycerol-coated (G) surface  for emitters NV$_{\text{A}}$--NV$_{\text{D}}$.} \label{Fig2}
\end{figure}
We begin our investigations with characterizing the charge state of $\SI{33}{}$ near-surface NV centers by applying the protocol illustrated in Fig.\ref{Fig1}(b). The $\SI{2}{\micro\second}$ short $\SI{532}{\nano\meter}$ pulse (power $\approx \SI{400}{\micro\watt}$; all laser powers specified herein are measured at the entrance to the objective) has the purpose to initialize the NV center into NV$^{-}$, followed by a weak $\SI{10}{\milli\second}$ long $\SI{594}{\nano\meter}$ pulse (power $\approx \SI{3.5}{\micro\watt}$) to selectively read out the charge state. The $\SI{594}{\nano \meter}$ readout laser mainly excites NV$^{-}$ as illustrated on the energy level diagrams in Fig.\ref{Fig1}(c), which leads to a charge state dependent fluorescence intensity. During the readout the NV center can ionize from NV$^{-}$ to NV$^{0}$ and recombine from NV$^{0}$ to NV$^{-}$ with the rates $\Gamma^{-0}$ and $\Gamma^{0-}$ respectively. The charge-state switching rates increase with increasing optical excitation power. The stability of the normalized fluorescence intensity during a readout pulse for NV$_{\text{A}}$ is shown in Fig.\ref{Fig1}(d). The photon-number distribution $P(n)$ obtained from NV$_{\text{A}}$ with the pristine diamond surface is plotted in Fig.~\ref{Fig1}(e), where the gray- and red-shaded areas illustrate the contributions of NV$^0$ and NV$^-$, respectively. We fitted the temporal evolution of $P(n)$ with a model~\cite{Bluvstein2019,Shields2015, Hac2018} ($\SI{100}{\micro\second}$ resolution) to extract the photon detection rates $\gamma^0$ and $\gamma^-$ from NV$^0$ and NV$^-$, respectively, $\Gamma^{-0}$, $\Gamma^{0-}$, and the NV$^-$ population P$_0$(NV$^{-}$) at the start of the readout pulse. With the pristine diamond surface, we obtain P$_0$(NV$^{-}$) = $\SI{0.56}{}$ and $\Gamma^{-0}=\SI{101}{\hertz}$, while $\Gamma^{0-}\approx\SI{6}{\hertz}$ (note that, due to a limited measurement time and a relatively high stability of the NV$^0$ charge state, it is not possible to quantify $\Gamma^{0-}$ with high accuracy within the measurement time). Afterwards, we coated the diamond surface with glycerol, resulting in a significant change of $P(n)$ as presented in Fig.\ref{Fig1}(f). The value of P$_0$(NV$^{-}$) substantially increases to 0.75 and is now comparable to results from NV centers deeper in the bulk~\cite{Aslam2013,Waldherr2011}. Furthermore, $\Gamma^{-0}$ substantially decreases to $\SI{9}{\hertz}$, demonstrating improved charge-state stability during readout after coating the diamond surface with glycerol.

The increased charge-state stability of NV$_{\text{A}}$ after application of glycerol is also directly seen in the average fluorescence during the readout pulse as shown in Fig.\ref{Fig1}(d). With the pristine surface, the fluorescence level reduces with a rate $\Gamma_{r,P} \approx \Gamma^{-0} + \Gamma^{0-} \approx \SI{107}{\hertz}$, while after application of glycerol the reduction is lower with a rate $\Gamma_{r,G} \approx \SI{25}{\hertz}$.

In Fig.~\ref{Fig2}(a), we plot the NV$^-$ initialization probability P$_0$(NV$^{-}$) versus the ionization rate $\Gamma^{-0}$ for $33$ single NV centers both with the pristine diamond surface and after the application of glycerol. The histogram for P$_0$(NV$^{-}$) is shown in Fig.~\ref{Fig2}(b).  We observe an anti-correlation between P$_0$(NV$^{-}$) and $\Gamma^{-0}$ irrespective of the surface condition. Upon the application of glycerol, P$_0$(NV$^{-}$) increases for all NV centers (see Fig.~\ref{Fig2}(c) for NV$_{\text{A}}$--NV$_{\text{D}}$) and correspondingly $\Gamma^{-0}$ reduces (see Fig.~\ref{Fig2}(d) for NV$_{\text{A}}$--NV$_{\text{D}}$).

Our results show that the application of glycerol generally increases the charge-initialization probability of the $\approx \SI{5}{\nano \meter}$ shallow NV centers into the NV$^{-}$ state with a maximum probability $\approx 75\%$ comparable to that of NV centers in the bulk. However, for some NV centers including NV$_{\text{B}}$, the charge-initialization probability (ionization rate) remains low (high) and only marginally improves compared to the pristine surface. 

\begin{figure}[!t]
\centering
\includegraphics [width=1\columnwidth]{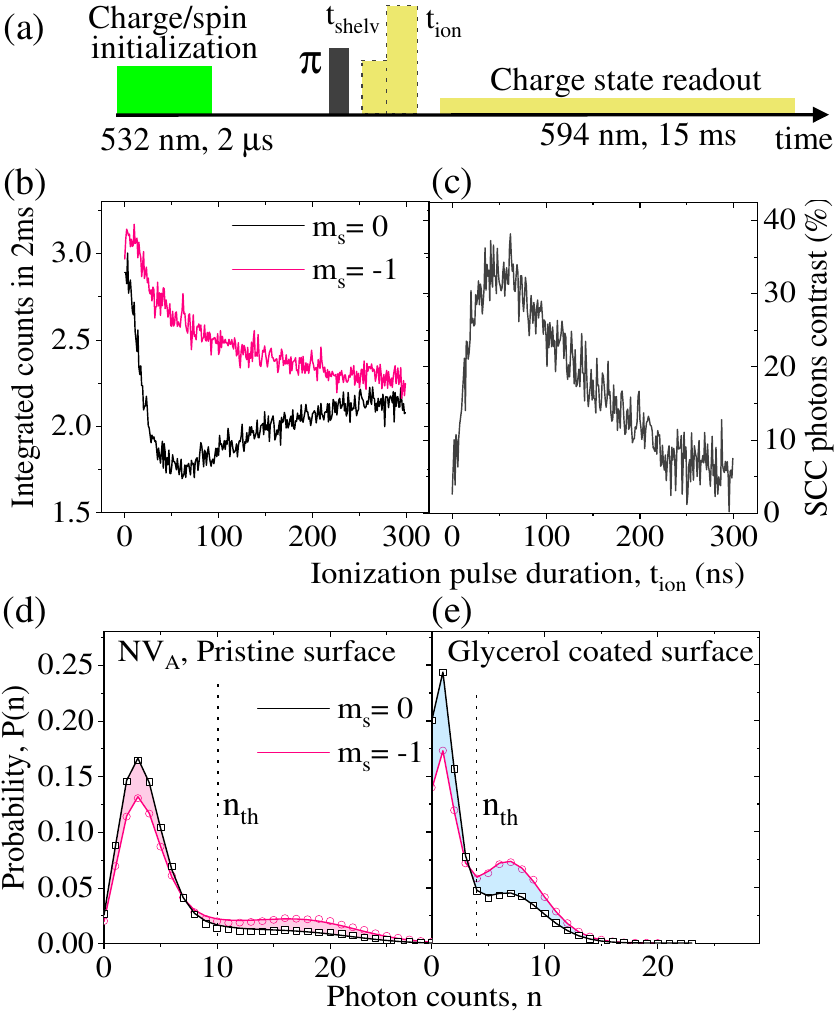}
\caption{SCC protocol and effects of surface condition. (a) Illustration of the pulse sequence to implement the SCC protocol, where the height of the orange ($\SI{594}{\nano \meter}$) pulses indicates the strength. (b) Photon counts integrated in $\SI{2}{\milli\second}$ of the readout pulse for initial $m_{s}=0$ and $m_{s}=-1$ spin states, respectively, and (c) contrast as a function of ionization-pulse duration $t_{ion}$. (d, e) The photon number distributions $P(n)$ of NV$_{\text{A}}$ for the initial spin states $m_s=0$ and $m_s=-1$ with the pristine (d) and the glycerol-coated surface (e). The optimal threshold photon number $n_{th}$ for the distributions is indicated.} 
\label{Fig3}
\end{figure}

\subsection{SCC protocol}
In the next step, we implement the SCC protocol [Fig.~\ref{Fig3}(a)] on four NV centers (NV$_{\text{A}}$--NV$_{\text{D}}$) that have different P$_0$(NV$^{-}$) and we benchmark the performance of SCC by estimating the spin-readout noise $\sigma_R$~\cite{Shields2015, Hopper2018a}.
Besides initialization into NV$^{-}$, the initial $\SI{2}{\micro\second}$, $\SI{532}{\nano\meter}$ pulse (power $\approx\SI{400}{\micro\watt}$) also prepares the spin state into m$_{s}$ = $\num{0}$, that subsequently can be transferred to m$_{s}$ =  $\num{-1}$ with a $\pi$ pulse. The following short $\SI{594}{\nano\meter}$ pulse (power $\approx\SI{4.4}{\milli\watt}$) then likely shelves the NV$^{-}$ center into the singlet state when in m$_{s}$ = $\num{-1}$ or leaves the population in the triplet state with a higher probability when in m$_{s}$ = $\num{0}$. Applied immediately after the shelving pulse, the triplet state of the NV center is ionized with a $\SI{29}{\milli\watt}$ (maximum available), $\SI{594}{\nano\meter}$ pulse of duration $t_{ion}$, thus effectively mapping the NV center spin states onto the distribution of charge states. Finally, the charge state is read out using a weak $\SI{15}{\milli\second}$ long (power $\SI{3.5}{\micro\watt}$) $\SI{594}{\nano\meter}$ pulse. The optimal duration and power of each pulse has been determined separately (see Supporting Information). As an example, in Fig.~\ref{Fig3}(b) we plot the integrated photon counts $\alpha_{m_s}$ acquired during the first $\SI{2}{\milli\second}$ readout window for the initial spin states $m_s=0$ and $m_s=-1$ as a function of the ionization-pulse duration $t_{ion}$. The SCC readout contrast $C = 1-\alpha_{0}/\alpha_{-1}$ is plotted in Fig.~\ref{Fig3}(c). The maximum contrast $C_{max} \approx 32 \%$ is found with an optimal $t_{ion} \approx \SI{50}{\nano\second}$. Similar to the mechanism of standard photoluminescence-based (PL) readout (see Supporting Information), the achievable contrast in SCC is limited by the spin-dependent intersystem crossing rates (see Supporting Information). With further increasing $t_{ion}$, the spin contrast decreases due to the relaxation of the singlet-state population back to the ground state and vanishes at $t_{ion} \approx \SI{300}{\nano\second}$, the approximate singlet-state lifetime. The spin-readout contrast with the glycerol-coated surface is similar to the value obtained with the pristine surface condition.

\begin{figure*}[ht!]
\centering
\includegraphics[width=\textwidth]{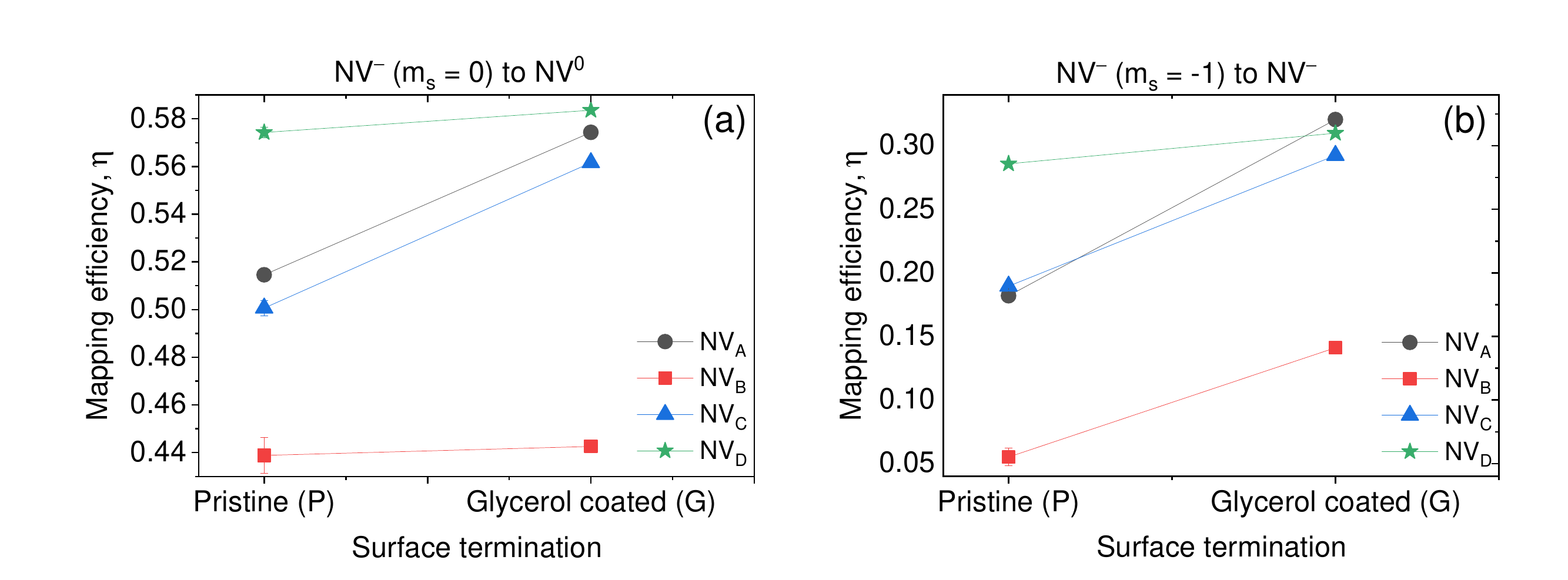}
\caption{Effect of diamond surface termination on the efficiency of spin-state dependent charge state mapping. The mapping efficiency for $m_S=0$ to NV$^{0}$ (a), and $m_S=-1$ to NV$^{-}$ (b). Glycerol coating improves the spin-to-charge conversion efficiency irrespective of the initial spin state of the NV$^{-}$ center.}
\label{Fig4}
\end{figure*}

The $m_s$-dependent photon-number distributions for NV$_{\text{A}}$ are plotted in Fig.~\ref{Fig3}(d) and~(e) for the pristine (d) and the glycerol-coated (e) surface, from which we extracted the respective NV$^-$ population $\tilde{p}_{m_s}$~\cite{Shields2015, Bluvstein2019}. While for the pristine surface, the difference $\tilde{p}_{-1} - \tilde{p}_{0} \approx 0.145$, the value increases to $\approx 0.19$ after adding glycerol, that is a direct result of the charge-state stability improved with glycerol.  
However, in the SCC-based spin readout, the final charge state is determined by counting photons and assigning the charge/spin states based on a threshold photon number $n_{th}$~\cite{Hopper2018a}. The precision of assigning the correct charge and thus spin state depends on the distinguishability of the photon number distributions $P_{m_s}(n)$ for the initial spin states $m_s$. In order to quantify the improved distinguishability of $P_{m_s}(n)$ after coating the surface with glycerol, we performed a Kolmogorov-Smirnov test that measures the maximum difference between the $m_s$ dependent cumulative photon distributions. The test yields values of 0.145 for the pristine surface and 0.174 for the glycerol-coated surface, an improvement of $\approx$ 20 $\%$. We note that no post-selection of the data was applied to discard for example events executed in the NV$^{0}$ state~\cite{Jaskula2019}.

The improvement in distinguishability arises due to better mapping of the spin-states onto the charges state. We quantify the mapping efficiency as the population transfer from the initial spin state, which is P$_0$(NV$^{-}$) assuming an ideal $\pi$ pulse, to the population in the charge states $\tilde{p}_{-,0}$:

\begin{equation}
\eta = \frac{\tilde{p}_{-,0}-\text{P}_0(\text{NV}^{-})}{\text{P}_0(\text{NV}^{-})}     
\end{equation}

The mapping efficiencies for NV$_{\text{A}}$ to NV$_{\text{D}}$ are shown in Fig.~\ref{Fig4}. For mapping  $m_s=0$ to NV$^0$, a modest improvement is seen when applying glycerol (Fig.~\ref{Fig4}a). The mapping efficiency for $m_S=-1$ to NV$^{-}$ (Fig.~\ref{Fig4}b) increases drastically when glycerol is applied to the surface and is around ~50\% for NV$_{\text{A}}$. The difference in improving the mapping efficiency for different spin states is not obvious, and understanding of this observation requires further investigation that involves time dynamics and transition rates between various states. One can also see from the plots in Fig. 4 that for NV$_{\text{D}}$, which was already stable with the pristine surface, addition of glycerol does not change much its charge-state mapping, while the efficiency for emitters with unstable charge environment benefits from surface passivation. Thus different NV centers might experience different effect of adding glycerol coating, which is eventually determined by the overall charge environment of the center.

\begin{figure}[htbp]
\centering
\includegraphics [width=1\columnwidth]{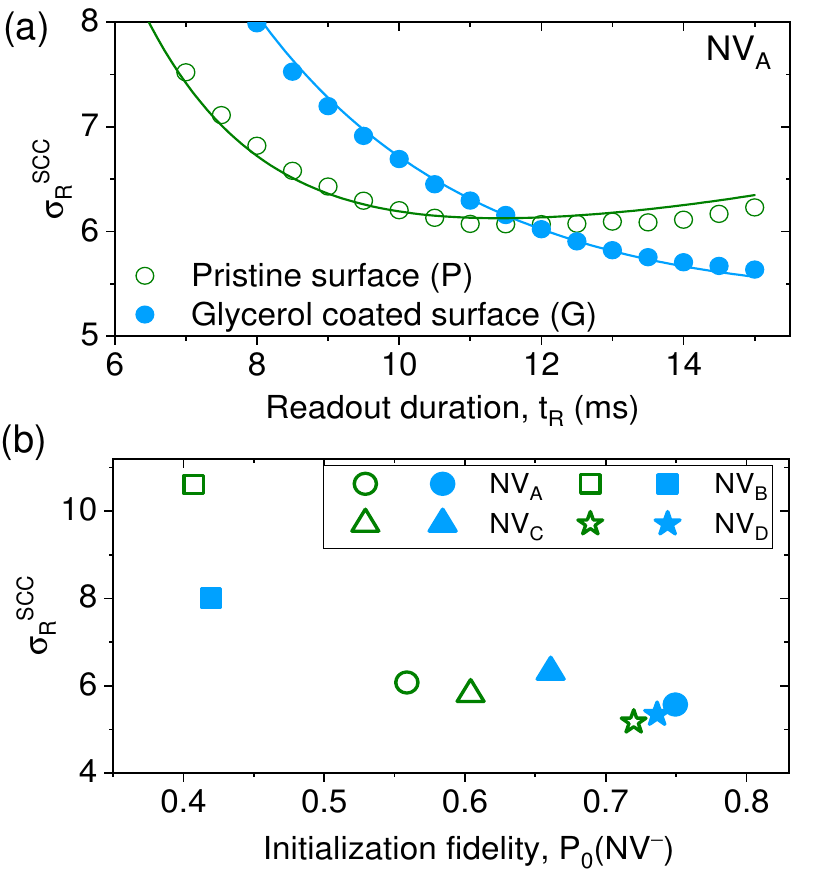}
\caption{Charge-initialization probability and spin-readout noise. (a) Measured spin-readout noise, $\sigma_{R}^{\text{SCC}}$ for NV$_{\text{A}}$ versus readout duration $t_R$. The symbols are calculated from the measured distributions [cf. Fig.~\ref{Fig3}(d) and (e)] and the solid lines are calculated from the extracted rates. (b) Optimal $\sigma_{R}^{\text{SCC}}$ obtained within $t_R < \SI{15}{\milli \second}$ readout time versus charge-initialization probability P$_0$(NV$^{-}$) for NV$_{\text{A}}$--NV$_{\text{D}}$. The green and blue symbols represent measured values with the pristine and the glycerol-coated surface, respectively. The impact of glycerol coating on $\sigma_{R}^{\text{SCC}}$ is substantial for the charge-unstable NV center, NV$_{\text{B}}$, while others show no improvement within $t_R < \SI{15}{\milli \second} $as a result of lower photon counts. The optimal values of $\sigma_R^{\text{SCC}}$ together with corresponding $t_R$ values obtained via simulation of the readout process for the glycerol-coated surface are summarized in Table~\ref{Tab:1}.}
\label{Fig5}
\end{figure}

For a set $n_{th}$ the spin-readout noise can be calculated via~\cite{Shields2015}
\begin{equation}
\sigma_{R}^{\text{SCC}}(t_{R}) = \sqrt{(p_{0}+p_{-1})(2-p_{0}-p_{-1})/(p_{0}-p_{-1})^2},
\end{equation}
where $p_{0}$ and $p_{-1}$ are the probabilities for the initial spin states $m_{s}=0$ and $m_{s}=-1$ to measure NV$^{-}$ at time $t_R$ during the readout step, respectively. Extracted from the distributions $P_{m_s}(n)$, in Fig.~\ref{Fig5}(a) we plot $\sigma_{R}^{\text{SCC}}(t_R)$ for NV$_{\text{A}}$ both for the pristine and the glycerol-coated surface. With increasing $t_R$, more photons are accumulated and the confidence for assigning the correct spin state improves. Hence, $\sigma_{R}^{\text{SCC}}(t_R)$ decreases to an optimal value. For longer $t_R$, charge-state switching occurs, thus causing uncertainty in determining the correct spin state and increasing $\sigma_{R}^{\text{SCC}}(t_R)$. With the glycerol-covered surface, $\sigma_{R}^{\text{SCC}}(t_R)$ is higher initially due to generally lower photon counts and the optimal $\sigma_{R}^{\text{SCC}}(t_R)$ is obtained at later times as a result of the improved charge-state stability. 

Fig. \ref{Fig5}(b) illustrates the optimal $\sigma_{R}^{\text{SCC}}(t_R)$ versus the charge initialization probability for NV$_{\text{A}}$--NV$_{\text{D}}$ obtained within a maximum readout time $t_R = \SI{15}{\milli\second}$. It is evident that the readout noise improves with increasing P$_0$(NV$^{-}$), respectively decreasing $\Gamma^{-0}$ (see Fig.~\ref{Fig2}(a)) irrespective of the surface condition.
The readout noise is higher for the charge-unstable NV centers (e.g. NV$_{\text{B}}$) and the improvement upon application of glycerol is significant. The influence of glycerol on $\sigma_{R}^{\text{SCC}}$ is minimal for the NV centers that are charge stable already with the pristine surface (e.g. NV$_{\text{C}}$ and NV$_{\text{D}}$). 
All key parameters for NV$_{\text{A}}$--NV$_{\text{D}}$ are summarized in Table.~\ref{Tab:1}, both for the pristine (P) and the glycerol-covered (G) surface. In the experiments, the maximum readout duration $t_R$ was $\SI{15}{\milli\second}$. However, the significantly reduced $\Gamma^{-0}$ after adding glycerol enables extended measurement times $t_R > \SI{15}{\milli\second}$. We thus simulated the measurement process (see Supporting Information) to extract the optimal $t_{R}$ and $n_{th}$ that minimizes $\sigma_{R}^{\text{SCC}}$ globally. It is clear, that for all four NV centers and due to  decreased $\Gamma^{-0}$, glycerol allows for extended measurement times $t_R > \SI{15}{\milli \second}$ as compared to the pristine surface. The minimum readout-noise we obtain is $\approx 5 \sigma_p$ at $t_R \approx \SI{12.8}{\milli \second}$  for NV$_{\text{D}}$ with P$_0$(NV$^{-}$) $\approx$ 0.75 and the pristine surface. This result is comparable to the readout noise previously obtained with a shallow NV center in a nanopillar~\cite{Ariyaratne2018} with significantly higher photon counts. We note that with the diamond sample turned over and the NV centers adjacent to the immersion oil with higher photon counts, we obtain a readout noise as low as $\approx 3.5 \sigma_p$. The relation between $\sigma_R$, $t_R$ and bandwidth normalized sensitivity will be addressed in the following section.

\begin{table*}[htbp]
\centering
\resizebox{\textwidth}{!}{
\begin{tabular}{|c|c|c|c|c|c|c|c|c|}
\hline
\rule{0pt}{3ex}   & P$_0$(NV$^{-}$) & $\Gamma^{-0}$ (Hz) & \thead{ \textbf{$\sigma_{R}^{\text{SCC}}$ ($\sigma_p$)} \\ (measured)} & \thead{\textbf{$\sigma_{R}^{\text{SCC}}$ ($\sigma_p$)} \\ (simulated)} & \thead{$n_{th}$ \\ (measured)} & \thead{$n_{th}$ \\ (simulated)} & \thead{$t_{R}$($ms$) \\ (measured)} & \thead{$t_{R}$($ms$) \\ (simulated)}\\ 

\hline
\rule{0pt}{3ex} &P\hspace{10mm}G  &P\hspace{10mm}G  &P\hspace{11mm}G &P\hspace{11mm}G &P\hspace{5mm}G &P\hspace{5mm}G &P\hspace{5mm}G &P\hspace{5mm}G\\ 
\hline
\rule{0pt}{3ex}NV$_{\text{A}}$ & 0.56\hspace{6mm}0.75 & 101\hspace{6mm}9 & 6.0\hspace{5mm}5.6 & 6.1\hspace{5mm}4.6 & 7\hspace{5mm}4 & 6\hspace{5mm}11 & 10.4\hspace{5mm}14.9 & 8.7\hspace{5mm}55.9\\ 

\rule{0pt}{3ex}NV$_{\text{B}}$ & 0.42\hspace{6mm}0.52 & 157\hspace{3mm}104 & 10.7\hspace{5mm}8.2 & 10.6\hspace{5mm}4.8 & 3\hspace{5mm}3 & 3\hspace{5mm}16 & 4.3\hspace{5mm}14.9 & 3.8\hspace{5mm}138\\ 

\rule{0pt}{3ex}NV$_{\text{C}}$ & 0.60\hspace{6mm}0.66 & 74\hspace{5mm}16 & 5.8\hspace{5mm}6.1 & 5.8\hspace{5mm}6.2 & 7\hspace{5mm}5 & 8\hspace{5mm}6 & 10.9\hspace{5mm}15 & 12.6\hspace{5mm}21.3\\ 

\rule{0pt}{3ex}NV$_{\text{D}}$ & 0.72\hspace{6mm}0.74 & 23\hspace{6mm}8 & 5.2\hspace{5mm}5.4 & 5.2\hspace{5mm}4.8 & 8\hspace{5mm}4 & 7\hspace{5mm}9 & 12.8\hspace{5mm}15 & 11.2\hspace{5mm}49.4\\ 
\hline
\end{tabular}
}
\caption{Summary of the key parameters measured and simulated for NV$_{\text{A}}$ - NV$_{\text{D}}$ for the pristine (P) and the glycerol-coated (G) surface condition. The uncertainty in $\sigma_R$ is around $0.1\sigma_p$.}
\label{Tab:1}
\end{table*}

\section{Discussion}
Deuterated glycerol not only increases the spin coherence of near-surface NV centers as shown previously~\cite{Kim2015}, but it generally improves the initialization probability P$_0$(NV$^{-}$) into the negative charge state and reduces the ionization rate $\Gamma^{-0}$ from NV$^-$ to NV$^0$. Our sample preparation routine with tri-acid cleaning only yields a partial surface termination with oxygen. Hence, the general improvement of the NV charge properties with glycerol could be assigned to the passivation of surface-related charges and traps ~\cite{Dhomkar2018, Bluvstein2019}. It is, however, also possible that glycerol lowers the Fermi level outside the diamond, reducing the effect of energy-band bending and thus improving the charge stability of NV$^-$~\cite{Hauf2011, Karaveli2016}. Irrespective of the physical mechanism, our findings highlight the influence of the surface-related charge environment on P$_0$(NV$^{-}$) as well as $\Gamma^{-0}$ during the readout with a weak selective $\SI{594}{\nano\meter}$ laser. 

In the following, we discuss the application of SCC in the context of a realistic sensing sequence and compared to the standard PL readout. The flat sample geometry, although practical, in our case limits the photon counts to approx. $\SI{65}{\kilo \, counts / \second}$ with $\SI{532}{\nano \meter}$ excitation at saturation. With PL spin readout in a $\SI{1}{\micro \second}$ window, we thus obtain $\sigma_{R}^{\text{PL}} \approx \num{30} \sigma_p$, a factor $\num{6}$ higher than the lowest $\sigma_{R}^{\text{SCC}} \approx 5.2 \sigma_p$ we obtained for NV$_{\text{D}}$. Our results demonstrate that even with a pristine surface, it is feasible to locate charge-stable NV centers, implanted $\approx \SI{5}{\nano \meter}$ deep, for which $\sigma_{R, min}^{\text{SCC}}$ is comparable to values previously only obtained from NV centers in nanobeams or deep inside bulk diamonds. However, the flat sample geometry as applied here and although practical, limits the attainable photon count rate. To minimize $\sigma_{R}^{\text{SCC}}$, the $t_{R}$ is thus required to be of the order of several $\si{\milli\second}$ to collect a sufficient number of photons. These optimal times are even longer with the glycerol-coated surface as the photon collection efficiency is further reduced. 
To account for the variability in the measurement overhead, we estimate the sensitivity to magnetic fields $\eta$ $\propto$ $\sigma_{R}(t_{R}) \sqrt{(\tau+t_{I}+t_{R})/\tau^{2}}$ \cite{Shields2015}, where $\tau$, $t_{I}$, and $t_{R}$ are the interrogation, the initialization and the readout times, respectively. Considering a similar initialization-pulse duration of $\SI{1}{\micro \second}$, and an optimal readout duration for PL ($\SI{1}{\micro \second}$) and SCC ($\SI{12}{\milli \second}$ for NV$_{\text{A}}$ with the pristine surface) minimizing $\sigma_{R}$, we obtain $\eta_{SCC} < \eta_{PL}$ for $\tau > \SI{500}{\micro \second}$. Despite the improved charge stability, for the glycerol covered surface the required $\tau$ would even be longer due to the further increased $t_R$ as a result of decreased photon counts as compared to the pristine surface. The increased distance to the potential sensing medium due to the glycerol layer thickness currently poses an additional challenge. 
For relaxometry measurements that are conducted on time scales T$_1 \approx \SI{1}{\milli \second}$, with our system parameters SCC thus would have an advantage compared to PL readout. The interrogation times $\tau > \SI{500}{\micro \second}$ required for SCC to be beneficial are much longer than the T$_2$ coherence times ($\numrange{3}{15} ~\SI{}{\micro \second}$) measured for the shallow NV centers in our sample. Increasing T$_2$ beyond $\SI{500}{\micro \second}$ by applying higher-order dynamical-decoupling sequences~\cite{Viola1999}, by coherently driving bath spins in the NV environment~\cite{Bluvstein2019a, Bauch2018}, or via application of a concatenated driving schemes~\cite{Cai2012, Stark2017} is challenging and remains to be explored experimentally. We thus anticipate that for the implementation of T$_{1}$-relaxometry sensing schemes on shallow NV centres with the pristine surface, it is advantageous to use the SCC instead of the PL readout. The minimum interrogation time beyond which SCC is beneficial can be further decreased by improving the photon collection via application of suitable photonic structures ~\cite{Hedrich2020, Wildanger2012} increasing the photon count rate.

The spin-to-charge readout may also be applied in the contex of quantum information processing with NV centers in diamonds at room temperature. In combination with a photoelectric readout, the spin-to-charge mapping with charge state readout has the potential of NV centre spin state readout with high spatial resolution~\cite{Siyushev2019}. Here a figure of merit is the single spin readout fidelity $\mathcal{F}$, which should be $>0.79$ for single-shot readout with SNR $>1$.\cite{Hopper2018a}. In our experiment, for NV$_{\text{D}}$ the readout fidelity is $\sim\!0.58$ and $\sim\!0.61$ with the pristine and glycerol-coated surface, respectively.
 
\section{Conclusion and Outlook}
In summary, in this article we investigated the charge-state stability of single NV centers implanted with a mean depth $\approx \SI{5}{\nano\meter}$ below the surface of a flat diamond sample. Applying standard processing and surface preparation by means of tri-acid cleaning~\cite{Brown2019}, we observe a strong variation in the probability for initializing the NV centers into the negative charge state. The NV$^-$ charge-state initialization probability increases for all investigated NV centers after coating the diamond surface with glycerol, and reaches up to $\approx 75\%$ that is comparable to NV centers deep in the bulk. Our observations thus demonstrate that deuterated glycerol does not only enhance the coherence time of near-surface NV centers~\cite{Kim2015}, but also improves the initialization probability and stability of the negative-charge state NV$^-$. While the role of surface electron traps on charge-state stability ~\cite{Dhomkar2018, Bluvstein2019,Stacey2019} and their detrimental effects on spin-based measurement protocols are recognized~\cite{Giri2018, Bluvstein2019}, the exact physical mechanism of our observations is not entirely clear. We hypothesize the observations may also be explained by a reduction of the effect of energy-band bending via a change of the local electric-field potential. Further investigations are thus required, including surface electron spectroscopy~\cite{Sangtawesin2019}, electrical manipulation via application of a gate voltage~\cite{Grotz2012}, and variation of different oxygen and hydrogen surface terminations~\cite{Hauf2011}.

We demonstrate the SCC protocol for shallow NV centers, obtaining a minimum readout-noise figure $\sigma_R^{SCC}$ in the range $\numrange{5}{6} \sigma_p$, a substantial reduction compared to standard PL readout with $\sigma_R^{PL} \approx 35 \sigma_p$. However, the reduction in $\sigma_R$ is achieved at the cost of substantially increased readout times. Previously, SCC has only been achieved with NV centers located far from the diamond surface where surface-related charge instabilities are insignificant or in nano-beams with higher photon collection efficiency. For NV centers where glycerol significantly lowers the ionization rate, thus improving charge stability during readout, we observe an improvement of $\sigma_R^{SCC}$ as prolonged readout times are allowed. However, the low photon count rate, limited by our sample configuration, requires extended readout times with the optimum $t_R > \SI{4}{\milli \second}$. With our configuration presented here, an improvement in terms of sensitivity is thus achieved for measurement sequences with an interrogation time $\tau > \SI{500}{\micro \second}$,  currently comprising relaxometry. The achievable interrogation times with dynamical decoupling and concatenated driving schemes remain to be explored. We expect that the optimal readout times can substantially be lowered by increasing the photon collection efficiency, reducing $t_R$ and yielding a benefit of SCC for interrogation times $< \SI{500}{\micro \second}$.

We envision our findings will motivate further investigations on the surface functionalization and charge state dynamics of shallow NV centres and potentially the application of the SCC protocol for near-surface NV centers with improved noise figures compared to standard PL readout. Structuring the diamond with for example nano-pillars will enhance the photon collection efficiency, thus improving the sensitivity even further for shorter interrogation times. Finally, we would like to note that although the long readout times, SCC can be the enabling technique such as in recently demonstrated covariance magnetometry~\cite{Rovny2022}.

\section{Acknowledgments}
We acknowledge financial support from the Novo Nordisk Foundation through the projects bioQ, bio-mag, and QuantBioEng, the Danish National Research Foundation (DNRF) through the center for Macroscopic Quantum States (bigQ, Grant No. DNRF0142), the Villum Foundation through the project bioCompass, and the EMPIR program co-financed by the Participating States and from the European Union’s Horizon 2020 research and innovation programme via the QADeT project (Grant No.20IND05).

%


\begin{thebibliography}{44}%
\makeatletter
\providecommand \@ifxundefined [1]{%
 \@ifx{#1\undefined}
}%
\providecommand \@ifnum [1]{%
 \ifnum #1\expandafter \@firstoftwo
 \else \expandafter \@secondoftwo
 \fi
}%
\providecommand \@ifx [1]{%
 \ifx #1\expandafter \@firstoftwo
 \else \expandafter \@secondoftwo
 \fi
}%
\providecommand \natexlab [1]{#1}%
\providecommand \enquote  [1]{``#1''}%
\providecommand \bibnamefont  [1]{#1}%
\providecommand \bibfnamefont [1]{#1}%
\providecommand \citenamefont [1]{#1}%
\providecommand \href@noop [0]{\@secondoftwo}%
\providecommand \href [0]{\begingroup \@sanitize@url \@href}%
\providecommand \@href[1]{\@@startlink{#1}\@@href}%
\providecommand \@@href[1]{\endgroup#1\@@endlink}%
\providecommand \@sanitize@url [0]{\catcode `\\12\catcode `\$12\catcode
  `\&12\catcode `\#12\catcode `\^12\catcode `\_12\catcode `\%12\relax}%
\providecommand \@@startlink[1]{}%
\providecommand \@@endlink[0]{}%
\providecommand \url  [0]{\begingroup\@sanitize@url \@url }%
\providecommand \@url [1]{\endgroup\@href {#1}{\urlprefix }}%
\providecommand \urlprefix  [0]{URL }%
\providecommand \Eprint [0]{\href }%
\providecommand \doibase [0]{https://doi.org/}%
\providecommand \selectlanguage [0]{\@gobble}%
\providecommand \bibinfo  [0]{\@secondoftwo}%
\providecommand \bibfield  [0]{\@secondoftwo}%
\providecommand \translation [1]{[#1]}%
\providecommand \BibitemOpen [0]{}%
\providecommand \bibitemStop [0]{}%
\providecommand \bibitemNoStop [0]{.\EOS\space}%
\providecommand \EOS [0]{\spacefactor3000\relax}%
\providecommand \BibitemShut  [1]{\csname bibitem#1\endcsname}%
\let\auto@bib@innerbib\@empty
\bibitem [{\citenamefont {Taylor}\ \emph {et~al.}(2008)\citenamefont {Taylor},
  \citenamefont {Cappellaro}, \citenamefont {Childress}, \citenamefont {Jiang},
  \citenamefont {Budker}, \citenamefont {Hemmer}, \citenamefont {Yacoby},
  \citenamefont {Walsworth},\ and\ \citenamefont {Lukin}}]{Taylor2008}%
  \BibitemOpen
  \bibfield  {author} {\bibinfo {author} {\bibfnamefont {J.~M.}\ \bibnamefont
  {Taylor}}, \bibinfo {author} {\bibfnamefont {P.}~\bibnamefont {Cappellaro}},
  \bibinfo {author} {\bibfnamefont {L.}~\bibnamefont {Childress}}, \bibinfo
  {author} {\bibfnamefont {L.}~\bibnamefont {Jiang}}, \bibinfo {author}
  {\bibfnamefont {D.}~\bibnamefont {Budker}}, \bibinfo {author} {\bibfnamefont
  {P.~R.}\ \bibnamefont {Hemmer}}, \bibinfo {author} {\bibfnamefont
  {A.}~\bibnamefont {Yacoby}}, \bibinfo {author} {\bibfnamefont
  {R.}~\bibnamefont {Walsworth}},\ and\ \bibinfo {author} {\bibfnamefont
  {M.~D.}\ \bibnamefont {Lukin}},\ }\bibfield  {title} {\bibinfo {title}
  {High-sensitivity diamond magnetometer with nanoscale resolution},\ }\href
  {https://doi.org/10.1038/nphys1075} {\bibfield  {journal} {\bibinfo
  {journal} {Nature Physics}\ }\textbf {\bibinfo {volume} {4}},\ \bibinfo
  {pages} {810} (\bibinfo {year} {2008})}\BibitemShut {NoStop}%
\bibitem [{\citenamefont {Balasubramanian}\ \emph {et~al.}(2008)\citenamefont
  {Balasubramanian}, \citenamefont {Chan}, \citenamefont {Kolesov},
  \citenamefont {Al-Hmoud}, \citenamefont {Tisler}, \citenamefont {Shin},
  \citenamefont {Kim}, \citenamefont {Wojcik}, \citenamefont {Hemmer},
  \citenamefont {Krueger}, \citenamefont {Hanke}, \citenamefont
  {Leitenstorfer}, \citenamefont {Bratschitsch}, \citenamefont {Jelezko},\ and\
  \citenamefont {Wrachtrup}}]{Balasubramanian2008}%
  \BibitemOpen
  \bibfield  {author} {\bibinfo {author} {\bibfnamefont {G.}~\bibnamefont
  {Balasubramanian}}, \bibinfo {author} {\bibfnamefont {I.~Y.}\ \bibnamefont
  {Chan}}, \bibinfo {author} {\bibfnamefont {R.}~\bibnamefont {Kolesov}},
  \bibinfo {author} {\bibfnamefont {M.}~\bibnamefont {Al-Hmoud}}, \bibinfo
  {author} {\bibfnamefont {J.}~\bibnamefont {Tisler}}, \bibinfo {author}
  {\bibfnamefont {C.}~\bibnamefont {Shin}}, \bibinfo {author} {\bibfnamefont
  {C.}~\bibnamefont {Kim}}, \bibinfo {author} {\bibfnamefont {A.}~\bibnamefont
  {Wojcik}}, \bibinfo {author} {\bibfnamefont {P.~R.}\ \bibnamefont {Hemmer}},
  \bibinfo {author} {\bibfnamefont {A.}~\bibnamefont {Krueger}}, \bibinfo
  {author} {\bibfnamefont {T.}~\bibnamefont {Hanke}}, \bibinfo {author}
  {\bibfnamefont {A.}~\bibnamefont {Leitenstorfer}}, \bibinfo {author}
  {\bibfnamefont {R.}~\bibnamefont {Bratschitsch}}, \bibinfo {author}
  {\bibfnamefont {F.}~\bibnamefont {Jelezko}},\ and\ \bibinfo {author}
  {\bibfnamefont {J.}~\bibnamefont {Wrachtrup}},\ }\bibfield  {title} {\bibinfo
  {title} {Nanoscale imaging magnetometry with diamond spins under ambient
  conditions},\ }\href {https://doi.org/10.1038/nature07278} {\bibfield
  {journal} {\bibinfo  {journal} {Nature}\ }\textbf {\bibinfo {volume} {455}},\
  \bibinfo {pages} {648} (\bibinfo {year} {2008})}\BibitemShut {NoStop}%
\bibitem [{\citenamefont {Dolde}\ \emph {et~al.}(2011)\citenamefont {Dolde},
  \citenamefont {Fedder}, \citenamefont {Doherty}, \citenamefont {Nöbauer},
  \citenamefont {Rempp}, \citenamefont {Balasubramanian}, \citenamefont {Wolf},
  \citenamefont {Reinhard}, \citenamefont {Hollenberg}, \citenamefont
  {Jelezko},\ and\ \citenamefont {Wrachtrup}}]{Dolde2011}%
  \BibitemOpen
  \bibfield  {author} {\bibinfo {author} {\bibfnamefont {F.}~\bibnamefont
  {Dolde}}, \bibinfo {author} {\bibfnamefont {H.}~\bibnamefont {Fedder}},
  \bibinfo {author} {\bibfnamefont {M.~W.}\ \bibnamefont {Doherty}}, \bibinfo
  {author} {\bibfnamefont {T.}~\bibnamefont {Nöbauer}}, \bibinfo {author}
  {\bibfnamefont {F.}~\bibnamefont {Rempp}}, \bibinfo {author} {\bibfnamefont
  {G.}~\bibnamefont {Balasubramanian}}, \bibinfo {author} {\bibfnamefont
  {T.}~\bibnamefont {Wolf}}, \bibinfo {author} {\bibfnamefont {F.}~\bibnamefont
  {Reinhard}}, \bibinfo {author} {\bibfnamefont {L.~C.~L.}\ \bibnamefont
  {Hollenberg}}, \bibinfo {author} {\bibfnamefont {F.}~\bibnamefont
  {Jelezko}},\ and\ \bibinfo {author} {\bibfnamefont {J.}~\bibnamefont
  {Wrachtrup}},\ }\bibfield  {title} {\bibinfo {title} {Electric-field sensing
  using single diamond spins},\ }\href {https://doi.org/10.1038/nphys1969}
  {\bibfield  {journal} {\bibinfo  {journal} {Nature Physics}\ }\textbf
  {\bibinfo {volume} {7}},\ \bibinfo {pages} {459} (\bibinfo {year}
  {2011})}\BibitemShut {NoStop}%
\bibitem [{\citenamefont {Doherty}\ \emph {et~al.}(2013)\citenamefont
  {Doherty}, \citenamefont {Manson}, \citenamefont {Delaney}, \citenamefont
  {Jelezko}, \citenamefont {Wrachtrup},\ and\ \citenamefont
  {Hollenberg}}]{Doherty2013}%
  \BibitemOpen
  \bibfield  {author} {\bibinfo {author} {\bibfnamefont {M.~W.}\ \bibnamefont
  {Doherty}}, \bibinfo {author} {\bibfnamefont {N.~B.}\ \bibnamefont {Manson}},
  \bibinfo {author} {\bibfnamefont {P.}~\bibnamefont {Delaney}}, \bibinfo
  {author} {\bibfnamefont {F.}~\bibnamefont {Jelezko}}, \bibinfo {author}
  {\bibfnamefont {J.}~\bibnamefont {Wrachtrup}},\ and\ \bibinfo {author}
  {\bibfnamefont {L.~C.}\ \bibnamefont {Hollenberg}},\ }\bibfield  {title}
  {\bibinfo {title} {The nitrogen-vacancy colour centre in diamond},\ }\href
  {https://doi.org/https://doi.org/10.1016/j.physrep.2013.02.001} {\bibfield
  {journal} {\bibinfo  {journal} {Physics Reports}\ }\textbf {\bibinfo {volume}
  {528}},\ \bibinfo {pages} {1} (\bibinfo {year} {2013})}\BibitemShut {NoStop}%
\bibitem [{\citenamefont {Steiner}\ \emph {et~al.}(2010)\citenamefont
  {Steiner}, \citenamefont {Neumann}, \citenamefont {Beck}, \citenamefont
  {Jelezko},\ and\ \citenamefont {Wrachtrup}}]{Steiner2010}%
  \BibitemOpen
  \bibfield  {author} {\bibinfo {author} {\bibfnamefont {M.}~\bibnamefont
  {Steiner}}, \bibinfo {author} {\bibfnamefont {P.}~\bibnamefont {Neumann}},
  \bibinfo {author} {\bibfnamefont {J.}~\bibnamefont {Beck}}, \bibinfo {author}
  {\bibfnamefont {F.}~\bibnamefont {Jelezko}},\ and\ \bibinfo {author}
  {\bibfnamefont {J.}~\bibnamefont {Wrachtrup}},\ }\bibfield  {title} {\bibinfo
  {title} {Universal enhancement of the optical readout fidelity of single
  electron spins at nitrogen-vacancy centers in diamond},\ }\href
  {https://doi.org/10.1103/PhysRevB.81.035205} {\bibfield  {journal} {\bibinfo
  {journal} {Phys. Rev. B}\ }\textbf {\bibinfo {volume} {81}},\ \bibinfo
  {pages} {035205} (\bibinfo {year} {2010})}\BibitemShut {NoStop}%
\bibitem [{\citenamefont {Jiang}\ \emph {et~al.}(2009)\citenamefont {Jiang},
  \citenamefont {Hodges}, \citenamefont {Maze}, \citenamefont {Maurer},
  \citenamefont {Taylor}, \citenamefont {Cory}, \citenamefont {Hemmer},
  \citenamefont {Walsworth}, \citenamefont {Yacoby}, \citenamefont {Zibrov},\
  and\ \citenamefont {Lukin}}]{Jiang2009}%
  \BibitemOpen
  \bibfield  {author} {\bibinfo {author} {\bibfnamefont {L.}~\bibnamefont
  {Jiang}}, \bibinfo {author} {\bibfnamefont {J.~S.}\ \bibnamefont {Hodges}},
  \bibinfo {author} {\bibfnamefont {J.~R.}\ \bibnamefont {Maze}}, \bibinfo
  {author} {\bibfnamefont {P.}~\bibnamefont {Maurer}}, \bibinfo {author}
  {\bibfnamefont {J.~M.}\ \bibnamefont {Taylor}}, \bibinfo {author}
  {\bibfnamefont {D.~G.}\ \bibnamefont {Cory}}, \bibinfo {author}
  {\bibfnamefont {P.~R.}\ \bibnamefont {Hemmer}}, \bibinfo {author}
  {\bibfnamefont {R.~L.}\ \bibnamefont {Walsworth}}, \bibinfo {author}
  {\bibfnamefont {A.}~\bibnamefont {Yacoby}}, \bibinfo {author} {\bibfnamefont
  {A.~S.}\ \bibnamefont {Zibrov}},\ and\ \bibinfo {author} {\bibfnamefont
  {M.~D.}\ \bibnamefont {Lukin}},\ }\bibfield  {title} {\bibinfo {title}
  {Repetitive readout of a single electronic spin via quantum logic with
  nuclear spin ancillae},\ }\href {https://doi.org/10.1126/science.1176496}
  {\bibfield  {journal} {\bibinfo  {journal} {Science}\ }\textbf {\bibinfo
  {volume} {326}},\ \bibinfo {pages} {267} (\bibinfo {year}
  {2009})}\BibitemShut {NoStop}%
\bibitem [{\citenamefont {Hedrich}\ \emph {et~al.}(2020)\citenamefont
  {Hedrich}, \citenamefont {Rohner}, \citenamefont {Batzer}, \citenamefont
  {Maletinsky},\ and\ \citenamefont {Shields}}]{Hedrich2020}%
  \BibitemOpen
  \bibfield  {author} {\bibinfo {author} {\bibfnamefont {N.}~\bibnamefont
  {Hedrich}}, \bibinfo {author} {\bibfnamefont {D.}~\bibnamefont {Rohner}},
  \bibinfo {author} {\bibfnamefont {M.}~\bibnamefont {Batzer}}, \bibinfo
  {author} {\bibfnamefont {P.}~\bibnamefont {Maletinsky}},\ and\ \bibinfo
  {author} {\bibfnamefont {B.~J.}\ \bibnamefont {Shields}},\ }\bibfield
  {title} {\bibinfo {title} {Parabolic diamond scanning probes for single-spin
  magnetic field imaging},\ }\href
  {https://doi.org/10.1103/PhysRevApplied.14.064007} {\bibfield  {journal}
  {\bibinfo  {journal} {Phys. Rev. Applied}\ }\textbf {\bibinfo {volume}
  {14}},\ \bibinfo {pages} {064007} (\bibinfo {year} {2020})}\BibitemShut
  {NoStop}%
\bibitem [{\citenamefont {Wildanger}\ \emph {et~al.}(2012)\citenamefont
  {Wildanger}, \citenamefont {Patton}, \citenamefont {Schill}, \citenamefont
  {Marseglia}, \citenamefont {Hadden}, \citenamefont {Knauer}, \citenamefont
  {Schönle}, \citenamefont {Rarity}, \citenamefont {O'Brien}, \citenamefont
  {Hell},\ and\ \citenamefont {Smith}}]{Wildanger2012}%
  \BibitemOpen
  \bibfield  {author} {\bibinfo {author} {\bibfnamefont {D.}~\bibnamefont
  {Wildanger}}, \bibinfo {author} {\bibfnamefont {B.~R.}\ \bibnamefont
  {Patton}}, \bibinfo {author} {\bibfnamefont {H.}~\bibnamefont {Schill}},
  \bibinfo {author} {\bibfnamefont {L.}~\bibnamefont {Marseglia}}, \bibinfo
  {author} {\bibfnamefont {J.~P.}\ \bibnamefont {Hadden}}, \bibinfo {author}
  {\bibfnamefont {S.}~\bibnamefont {Knauer}}, \bibinfo {author} {\bibfnamefont
  {A.}~\bibnamefont {Schönle}}, \bibinfo {author} {\bibfnamefont {J.~G.}\
  \bibnamefont {Rarity}}, \bibinfo {author} {\bibfnamefont {J.~L.}\
  \bibnamefont {O'Brien}}, \bibinfo {author} {\bibfnamefont {S.~W.}\
  \bibnamefont {Hell}},\ and\ \bibinfo {author} {\bibfnamefont {J.~M.}\
  \bibnamefont {Smith}},\ }\bibfield  {title} {\bibinfo {title} {Solid
  immersion facilitates fluorescence microscopy with nanometer resolution and
  sub-angstrom emitter localization},\ }\href
  {https://doi.org/https://doi.org/10.1002/adma.201203033} {\bibfield
  {journal} {\bibinfo  {journal} {Advanced Materials}\ }\textbf {\bibinfo
  {volume} {24}},\ \bibinfo {pages} {OP309} (\bibinfo {year}
  {2012})}\BibitemShut {NoStop}%
\bibitem [{\citenamefont {Siyushev}\ \emph {et~al.}(2019)\citenamefont
  {Siyushev}, \citenamefont {Nesladek}, \citenamefont {Bourgeois},
  \citenamefont {Gulka}, \citenamefont {Hruby}, \citenamefont {Yamamoto},
  \citenamefont {Trupke}, \citenamefont {Teraji}, \citenamefont {Isoya},\ and\
  \citenamefont {Jelezko}}]{Siyushev2019}%
  \BibitemOpen
  \bibfield  {author} {\bibinfo {author} {\bibfnamefont {P.}~\bibnamefont
  {Siyushev}}, \bibinfo {author} {\bibfnamefont {M.}~\bibnamefont {Nesladek}},
  \bibinfo {author} {\bibfnamefont {E.}~\bibnamefont {Bourgeois}}, \bibinfo
  {author} {\bibfnamefont {M.}~\bibnamefont {Gulka}}, \bibinfo {author}
  {\bibfnamefont {J.}~\bibnamefont {Hruby}}, \bibinfo {author} {\bibfnamefont
  {T.}~\bibnamefont {Yamamoto}}, \bibinfo {author} {\bibfnamefont
  {M.}~\bibnamefont {Trupke}}, \bibinfo {author} {\bibfnamefont
  {T.}~\bibnamefont {Teraji}}, \bibinfo {author} {\bibfnamefont
  {J.}~\bibnamefont {Isoya}},\ and\ \bibinfo {author} {\bibfnamefont
  {F.}~\bibnamefont {Jelezko}},\ }\bibfield  {title} {\bibinfo {title}
  {Photoelectrical imaging and coherent spin-state readout of single
  nitrogen-vacancy centers in diamond},\ }\href
  {https://doi.org/10.1126/science.aav2789} {\bibfield  {journal} {\bibinfo
  {journal} {Science}\ }\textbf {\bibinfo {volume} {363}},\ \bibinfo {pages}
  {728} (\bibinfo {year} {2019})},\ \Eprint
  {https://arxiv.org/abs/https://www.science.org/doi/pdf/10.1126/science.aav2789}
  {https://www.science.org/doi/pdf/10.1126/science.aav2789} \BibitemShut
  {NoStop}%
\bibitem [{\citenamefont {Dhomkar}\ \emph {et~al.}(2018)\citenamefont
  {Dhomkar}, \citenamefont {Jayakumar}, \citenamefont {Zangara},\ and\
  \citenamefont {Meriles}}]{Dhomkar2018}%
  \BibitemOpen
  \bibfield  {author} {\bibinfo {author} {\bibfnamefont {S.}~\bibnamefont
  {Dhomkar}}, \bibinfo {author} {\bibfnamefont {H.}~\bibnamefont {Jayakumar}},
  \bibinfo {author} {\bibfnamefont {P.~R.}\ \bibnamefont {Zangara}},\ and\
  \bibinfo {author} {\bibfnamefont {C.~A.}\ \bibnamefont {Meriles}},\
  }\bibfield  {title} {\bibinfo {title} {Charge dynamics in near-surface,
  variable-density ensembles of nitrogen-vacancy centers in diamond},\ }\href
  {https://doi.org/10.1021/acs.nanolett.8b01739} {\bibfield  {journal}
  {\bibinfo  {journal} {Nano Letters}\ }\textbf {\bibinfo {volume} {18}},\
  \bibinfo {pages} {4046} (\bibinfo {year} {2018})}\BibitemShut {NoStop}%
\bibitem [{\citenamefont {Bluvstein}\ \emph
  {et~al.}(2019{\natexlab{a}})\citenamefont {Bluvstein}, \citenamefont
  {Zhang},\ and\ \citenamefont {Jayich}}]{Bluvstein2019}%
  \BibitemOpen
  \bibfield  {author} {\bibinfo {author} {\bibfnamefont {D.}~\bibnamefont
  {Bluvstein}}, \bibinfo {author} {\bibfnamefont {Z.}~\bibnamefont {Zhang}},\
  and\ \bibinfo {author} {\bibfnamefont {A.~C.~B.}\ \bibnamefont {Jayich}},\
  }\bibfield  {title} {\bibinfo {title} {Identifying and mitigating charge
  instabilities in shallow diamond nitrogen-vacancy centers},\ }\href
  {https://doi.org/10.1103/PhysRevLett.122.076101} {\bibfield  {journal}
  {\bibinfo  {journal} {Phys Rev Lett}\ }\textbf {\bibinfo {volume} {122}},\
  \bibinfo {pages} {076101} (\bibinfo {year} {2019}{\natexlab{a}})}\BibitemShut
  {NoStop}%
\bibitem [{\citenamefont {Gorrini}\ \emph {et~al.}(2021)\citenamefont
  {Gorrini}, \citenamefont {Dorigoni}, \citenamefont {Olivares-Postigo},
  \citenamefont {Giri}, \citenamefont {Aprà}, \citenamefont {Picollo},\ and\
  \citenamefont {Bifone}}]{Gorrini2021}%
  \BibitemOpen
  \bibfield  {author} {\bibinfo {author} {\bibfnamefont {F.}~\bibnamefont
  {Gorrini}}, \bibinfo {author} {\bibfnamefont {C.}~\bibnamefont {Dorigoni}},
  \bibinfo {author} {\bibfnamefont {D.}~\bibnamefont {Olivares-Postigo}},
  \bibinfo {author} {\bibfnamefont {R.}~\bibnamefont {Giri}}, \bibinfo {author}
  {\bibfnamefont {P.}~\bibnamefont {Aprà}}, \bibinfo {author} {\bibfnamefont
  {F.}~\bibnamefont {Picollo}},\ and\ \bibinfo {author} {\bibfnamefont
  {A.}~\bibnamefont {Bifone}},\ }\bibfield  {title} {\bibinfo {title}
  {Long-lived ensembles of shallow nv$^{-}$ centers in flat and nanostructured
  diamonds by photoconversion},\ }\href
  {https://doi.org/10.1021/acsami.1c09825} {\bibfield  {journal} {\bibinfo
  {journal} {ACS Applied Materials \& Interfaces}\ }\textbf {\bibinfo {volume}
  {13}},\ \bibinfo {pages} {43221} (\bibinfo {year} {2021})}\BibitemShut
  {NoStop}%
\bibitem [{\citenamefont {Aslam}\ \emph {et~al.}(2013)\citenamefont {Aslam},
  \citenamefont {Waldherr}, \citenamefont {Neumann}, \citenamefont {Jelezko},\
  and\ \citenamefont {Wrachtrup}}]{Aslam2013}%
  \BibitemOpen
  \bibfield  {author} {\bibinfo {author} {\bibfnamefont {N.}~\bibnamefont
  {Aslam}}, \bibinfo {author} {\bibfnamefont {G.}~\bibnamefont {Waldherr}},
  \bibinfo {author} {\bibfnamefont {P.}~\bibnamefont {Neumann}}, \bibinfo
  {author} {\bibfnamefont {F.}~\bibnamefont {Jelezko}},\ and\ \bibinfo {author}
  {\bibfnamefont {J.}~\bibnamefont {Wrachtrup}},\ }\bibfield  {title} {\bibinfo
  {title} {Photo-induced ionization dynamics of the nitrogen vacancy defect in
  diamond investigated by single-shot charge state detection},\ }\href
  {https://doi.org/10.1088/1367-2630/15/1/013064} {\bibfield  {journal}
  {\bibinfo  {journal} {New Journal of Physics}\ }\textbf {\bibinfo {volume}
  {15}},\ \bibinfo {pages} {013064} (\bibinfo {year} {2013})}\BibitemShut
  {NoStop}%
\bibitem [{\citenamefont {Shields}\ \emph {et~al.}(2015)\citenamefont
  {Shields}, \citenamefont {Unterreithmeier}, \citenamefont {de~Leon},
  \citenamefont {Park},\ and\ \citenamefont {Lukin}}]{Shields2015}%
  \BibitemOpen
  \bibfield  {author} {\bibinfo {author} {\bibfnamefont {B.~J.}\ \bibnamefont
  {Shields}}, \bibinfo {author} {\bibfnamefont {Q.~P.}\ \bibnamefont
  {Unterreithmeier}}, \bibinfo {author} {\bibfnamefont {N.~P.}\ \bibnamefont
  {de~Leon}}, \bibinfo {author} {\bibfnamefont {H.}~\bibnamefont {Park}},\ and\
  \bibinfo {author} {\bibfnamefont {M.~D.}\ \bibnamefont {Lukin}},\ }\bibfield
  {title} {\bibinfo {title} {Efficient readout of a single spin state in
  diamond via spin-to-charge conversion},\ }\href
  {https://doi.org/10.1103/PhysRevLett.114.136402} {\bibfield  {journal}
  {\bibinfo  {journal} {Phys Rev Lett}\ }\textbf {\bibinfo {volume} {114}},\
  \bibinfo {pages} {136402} (\bibinfo {year} {2015})}\BibitemShut {NoStop}%
\bibitem [{\citenamefont {Manson}\ and\ \citenamefont
  {Harrison}(2005)}]{Manson2005}%
  \BibitemOpen
  \bibfield  {author} {\bibinfo {author} {\bibfnamefont {N.}~\bibnamefont
  {Manson}}\ and\ \bibinfo {author} {\bibfnamefont {J.}~\bibnamefont
  {Harrison}},\ }\bibfield  {title} {\bibinfo {title} {Photo-ionization of the
  nitrogen-vacancy center in diamond},\ }\href
  {https://doi.org/https://doi.org/10.1016/j.diamond.2005.06.027} {\bibfield
  {journal} {\bibinfo  {journal} {Diamond and Related Materials}\ }\textbf
  {\bibinfo {volume} {14}},\ \bibinfo {pages} {1705} (\bibinfo {year}
  {2005})}\BibitemShut {NoStop}%
\bibitem [{\citenamefont {Hopper}\ \emph
  {et~al.}(2018{\natexlab{a}})\citenamefont {Hopper}, \citenamefont
  {Shulevitz},\ and\ \citenamefont {Bassett}}]{Hopper2018a}%
  \BibitemOpen
  \bibfield  {author} {\bibinfo {author} {\bibfnamefont {D.~A.}\ \bibnamefont
  {Hopper}}, \bibinfo {author} {\bibfnamefont {H.~J.}\ \bibnamefont
  {Shulevitz}},\ and\ \bibinfo {author} {\bibfnamefont {L.~C.}\ \bibnamefont
  {Bassett}},\ }\bibfield  {title} {\bibinfo {title} {Spin readout techniques
  of the nitrogen-vacancy center in diamond},\ }\href
  {https://doi.org/10.3390/mi9090437} {\bibfield  {journal} {\bibinfo
  {journal} {Micromachines (Basel)}\ }\textbf {\bibinfo {volume} {9}},\
  \bibinfo {pages} {437} (\bibinfo {year} {2018}{\natexlab{a}})}\BibitemShut
  {NoStop}%
\bibitem [{\citenamefont {Hopper}\ \emph
  {et~al.}(2018{\natexlab{b}})\citenamefont {Hopper}, \citenamefont {Grote},
  \citenamefont {Parks},\ and\ \citenamefont {Bassett}}]{Hopper2018}%
  \BibitemOpen
  \bibfield  {author} {\bibinfo {author} {\bibfnamefont {D.~A.}\ \bibnamefont
  {Hopper}}, \bibinfo {author} {\bibfnamefont {R.~R.}\ \bibnamefont {Grote}},
  \bibinfo {author} {\bibfnamefont {S.~M.}\ \bibnamefont {Parks}},\ and\
  \bibinfo {author} {\bibfnamefont {L.~C.}\ \bibnamefont {Bassett}},\
  }\bibfield  {title} {\bibinfo {title} {Amplified sensitivity of
  nitrogen-vacancy spins in nanodiamonds using all-optical charge readout},\
  }\href {https://doi.org/10.1021/acsnano.8b01265} {\bibfield  {journal}
  {\bibinfo  {journal} {ACS Nano}\ }\textbf {\bibinfo {volume} {12}},\ \bibinfo
  {pages} {4678} (\bibinfo {year} {2018}{\natexlab{b}})}\BibitemShut {NoStop}%
\bibitem [{\citenamefont {Jayakumar}\ \emph {et~al.}(2018)\citenamefont
  {Jayakumar}, \citenamefont {Dhomkar}, \citenamefont {Henshaw},\ and\
  \citenamefont {Meriles}}]{Jayakumar2018}%
  \BibitemOpen
  \bibfield  {author} {\bibinfo {author} {\bibfnamefont {H.}~\bibnamefont
  {Jayakumar}}, \bibinfo {author} {\bibfnamefont {S.}~\bibnamefont {Dhomkar}},
  \bibinfo {author} {\bibfnamefont {J.}~\bibnamefont {Henshaw}},\ and\ \bibinfo
  {author} {\bibfnamefont {C.~A.}\ \bibnamefont {Meriles}},\ }\bibfield
  {title} {\bibinfo {title} {Spin readout via spin-to-charge conversion in bulk
  diamond nitrogen-vacancy ensembles},\ }\href
  {https://doi.org/10.1063/1.5040261} {\bibfield  {journal} {\bibinfo
  {journal} {Applied Physics Letters}\ }\textbf {\bibinfo {volume} {113}},\
  \bibinfo {pages} {122404} (\bibinfo {year} {2018})}\BibitemShut {NoStop}%
\bibitem [{\citenamefont {Jaskula}\ \emph {et~al.}(2019)\citenamefont
  {Jaskula}, \citenamefont {Shields}, \citenamefont {Bauch}, \citenamefont
  {Lukin}, \citenamefont {Trifonov},\ and\ \citenamefont
  {Walsworth}}]{Jaskula2019}%
  \BibitemOpen
  \bibfield  {author} {\bibinfo {author} {\bibfnamefont {J.~C.}\ \bibnamefont
  {Jaskula}}, \bibinfo {author} {\bibfnamefont {B.~J.}\ \bibnamefont
  {Shields}}, \bibinfo {author} {\bibfnamefont {E.}~\bibnamefont {Bauch}},
  \bibinfo {author} {\bibfnamefont {M.~D.}\ \bibnamefont {Lukin}}, \bibinfo
  {author} {\bibfnamefont {A.~S.}\ \bibnamefont {Trifonov}},\ and\ \bibinfo
  {author} {\bibfnamefont {R.~L.}\ \bibnamefont {Walsworth}},\ }\bibfield
  {title} {\bibinfo {title} {Improved quantum sensing with a single solid-state
  spin via spin-to-charge conversion},\ }\href
  {https://doi.org/10.1103/PhysRevApplied.11.064003} {\bibfield  {journal}
  {\bibinfo  {journal} {Physical Review Applied}\ }\textbf {\bibinfo {volume}
  {11}},\ \bibinfo {pages} {064003} (\bibinfo {year} {2019})}\BibitemShut
  {NoStop}%
\bibitem [{\citenamefont {Rovny}\ \emph {et~al.}(2022)\citenamefont {Rovny},
  \citenamefont {Yuan}, \citenamefont {Fitzpatrick}, \citenamefont {Abdalla},
  \citenamefont {Futamura}, \citenamefont {Fox}, \citenamefont {Cambria},
  \citenamefont {Kolkowitz},\ and\ \citenamefont {de~Leon}}]{Rovny2022}%
  \BibitemOpen
  \bibfield  {author} {\bibinfo {author} {\bibfnamefont {J.}~\bibnamefont
  {Rovny}}, \bibinfo {author} {\bibfnamefont {Z.}~\bibnamefont {Yuan}},
  \bibinfo {author} {\bibfnamefont {M.}~\bibnamefont {Fitzpatrick}}, \bibinfo
  {author} {\bibfnamefont {A.~I.}\ \bibnamefont {Abdalla}}, \bibinfo {author}
  {\bibfnamefont {L.}~\bibnamefont {Futamura}}, \bibinfo {author}
  {\bibfnamefont {C.}~\bibnamefont {Fox}}, \bibinfo {author} {\bibfnamefont
  {M.~C.}\ \bibnamefont {Cambria}}, \bibinfo {author} {\bibfnamefont
  {S.}~\bibnamefont {Kolkowitz}},\ and\ \bibinfo {author} {\bibfnamefont
  {N.~P.}\ \bibnamefont {de~Leon}},\ }\bibfield  {title} {\bibinfo {title}
  {Nanoscale covariance magnetometry with diamond quantum sensors},\ }\href
  {https://doi.org/10.1126/science.ade9858} {\bibfield  {journal} {\bibinfo
  {journal} {Science}\ }\textbf {\bibinfo {volume} {378}},\ \bibinfo {pages}
  {1301} (\bibinfo {year} {2022})},\ \Eprint
  {https://arxiv.org/abs/https://www.science.org/doi/pdf/10.1126/science.ade9858}
  {https://www.science.org/doi/pdf/10.1126/science.ade9858} \BibitemShut
  {NoStop}%
\bibitem [{\citenamefont {Mamin}\ \emph {et~al.}(2013)\citenamefont {Mamin},
  \citenamefont {Kim}, \citenamefont {Sherwood}, \citenamefont {Rettner},
  \citenamefont {Ohno}, \citenamefont {Awschalom},\ and\ \citenamefont
  {Rugar}}]{Mamin2013}%
  \BibitemOpen
  \bibfield  {author} {\bibinfo {author} {\bibfnamefont {H.~J.}\ \bibnamefont
  {Mamin}}, \bibinfo {author} {\bibfnamefont {M.}~\bibnamefont {Kim}}, \bibinfo
  {author} {\bibfnamefont {M.~H.}\ \bibnamefont {Sherwood}}, \bibinfo {author}
  {\bibfnamefont {C.~T.}\ \bibnamefont {Rettner}}, \bibinfo {author}
  {\bibfnamefont {K.}~\bibnamefont {Ohno}}, \bibinfo {author} {\bibfnamefont
  {D.~D.}\ \bibnamefont {Awschalom}},\ and\ \bibinfo {author} {\bibfnamefont
  {D.}~\bibnamefont {Rugar}},\ }\bibfield  {title} {\bibinfo {title} {Nanoscale
  nuclear magnetic resonance with a nitrogen-vacancy spin sensor},\ }\href
  {https://doi.org/10.1126/science.1231540} {\bibfield  {journal} {\bibinfo
  {journal} {Science}\ }\textbf {\bibinfo {volume} {339}},\ \bibinfo {pages}
  {557} (\bibinfo {year} {2013})}\BibitemShut {NoStop}%
\bibitem [{\citenamefont {Staudacher}\ \emph {et~al.}(2013)\citenamefont
  {Staudacher}, \citenamefont {Shi}, \citenamefont {Pezzagna}, \citenamefont
  {Meijer}, \citenamefont {Du}, \citenamefont {Meriles}, \citenamefont
  {Reinhard},\ and\ \citenamefont {Wrachtrup}}]{Staudacher2013}%
  \BibitemOpen
  \bibfield  {author} {\bibinfo {author} {\bibfnamefont {T.}~\bibnamefont
  {Staudacher}}, \bibinfo {author} {\bibfnamefont {F.}~\bibnamefont {Shi}},
  \bibinfo {author} {\bibfnamefont {S.}~\bibnamefont {Pezzagna}}, \bibinfo
  {author} {\bibfnamefont {J.}~\bibnamefont {Meijer}}, \bibinfo {author}
  {\bibfnamefont {J.}~\bibnamefont {Du}}, \bibinfo {author} {\bibfnamefont
  {C.~A.}\ \bibnamefont {Meriles}}, \bibinfo {author} {\bibfnamefont
  {F.}~\bibnamefont {Reinhard}},\ and\ \bibinfo {author} {\bibfnamefont
  {J.}~\bibnamefont {Wrachtrup}},\ }\bibfield  {title} {\bibinfo {title}
  {Nuclear magnetic resonance spectroscopy on a (5-nanometer)$^{3}$ sample
  volume},\ }\href {https://doi.org/10.1126/science.1231675} {\bibfield
  {journal} {\bibinfo  {journal} {Science}\ }\textbf {\bibinfo {volume}
  {339}},\ \bibinfo {pages} {561} (\bibinfo {year} {2013})}\BibitemShut
  {NoStop}%
\bibitem [{\citenamefont {Perona~Martinez}\ \emph {et~al.}(2020)\citenamefont
  {Perona~Martinez}, \citenamefont {Nusantara}, \citenamefont {Chipaux},
  \citenamefont {Padamati},\ and\ \citenamefont {Schirhagl}}]{Martinez2020}%
  \BibitemOpen
  \bibfield  {author} {\bibinfo {author} {\bibfnamefont {F.}~\bibnamefont
  {Perona~Martinez}}, \bibinfo {author} {\bibfnamefont {A.~C.}\ \bibnamefont
  {Nusantara}}, \bibinfo {author} {\bibfnamefont {M.}~\bibnamefont {Chipaux}},
  \bibinfo {author} {\bibfnamefont {S.~K.}\ \bibnamefont {Padamati}},\ and\
  \bibinfo {author} {\bibfnamefont {R.}~\bibnamefont {Schirhagl}},\ }\bibfield
  {title} {\bibinfo {title} {Nanodiamond relaxometry-based detection of
  free-radical species when produced in chemical reactions in biologically
  relevant conditions},\ }\href {https://doi.org/10.1021/acssensors.0c01037}
  {\bibfield  {journal} {\bibinfo  {journal} {ACS Sensors}\ }\textbf {\bibinfo
  {volume} {5}},\ \bibinfo {pages} {3862} (\bibinfo {year} {2020})}\BibitemShut
  {NoStop}%
\bibitem [{\citenamefont {Muller}\ \emph {et~al.}(2014)\citenamefont {Muller},
  \citenamefont {Kong}, \citenamefont {Cai}, \citenamefont {Melentijevic},
  \citenamefont {Stacey}, \citenamefont {Markham}, \citenamefont {Twitchen},
  \citenamefont {Isoya}, \citenamefont {Pezzagna}, \citenamefont {Meijer},
  \citenamefont {Du}, \citenamefont {Plenio}, \citenamefont {Naydenov},
  \citenamefont {McGuinness},\ and\ \citenamefont {Jelezko}}]{Muller2014}%
  \BibitemOpen
  \bibfield  {author} {\bibinfo {author} {\bibfnamefont {C.}~\bibnamefont
  {Muller}}, \bibinfo {author} {\bibfnamefont {X.}~\bibnamefont {Kong}},
  \bibinfo {author} {\bibfnamefont {J.~M.}\ \bibnamefont {Cai}}, \bibinfo
  {author} {\bibfnamefont {K.}~\bibnamefont {Melentijevic}}, \bibinfo {author}
  {\bibfnamefont {A.}~\bibnamefont {Stacey}}, \bibinfo {author} {\bibfnamefont
  {M.}~\bibnamefont {Markham}}, \bibinfo {author} {\bibfnamefont
  {D.}~\bibnamefont {Twitchen}}, \bibinfo {author} {\bibfnamefont
  {J.}~\bibnamefont {Isoya}}, \bibinfo {author} {\bibfnamefont
  {S.}~\bibnamefont {Pezzagna}}, \bibinfo {author} {\bibfnamefont
  {J.}~\bibnamefont {Meijer}}, \bibinfo {author} {\bibfnamefont {J.~F.}\
  \bibnamefont {Du}}, \bibinfo {author} {\bibfnamefont {M.~B.}\ \bibnamefont
  {Plenio}}, \bibinfo {author} {\bibfnamefont {B.}~\bibnamefont {Naydenov}},
  \bibinfo {author} {\bibfnamefont {L.~P.}\ \bibnamefont {McGuinness}},\ and\
  \bibinfo {author} {\bibfnamefont {F.}~\bibnamefont {Jelezko}},\ }\bibfield
  {title} {\bibinfo {title} {Nuclear magnetic resonance spectroscopy with
  single spin sensitivity},\ }\href {https://doi.org/10.1038/ncomms5703}
  {\bibfield  {journal} {\bibinfo  {journal} {Nat Commun}\ }\textbf {\bibinfo
  {volume} {5}},\ \bibinfo {pages} {4703} (\bibinfo {year} {2014})}\BibitemShut
  {NoStop}%
\bibitem [{\citenamefont {Hauf}\ \emph {et~al.}(2011)\citenamefont {Hauf},
  \citenamefont {Grotz}, \citenamefont {Naydenov}, \citenamefont {Dankerl},
  \citenamefont {Pezzagna}, \citenamefont {Meijer}, \citenamefont {Jelezko},
  \citenamefont {Wrachtrup}, \citenamefont {Stutzmann}, \citenamefont
  {Reinhard},\ and\ \citenamefont {Garrido}}]{Hauf2011}%
  \BibitemOpen
  \bibfield  {author} {\bibinfo {author} {\bibfnamefont {M.~V.}\ \bibnamefont
  {Hauf}}, \bibinfo {author} {\bibfnamefont {B.}~\bibnamefont {Grotz}},
  \bibinfo {author} {\bibfnamefont {B.}~\bibnamefont {Naydenov}}, \bibinfo
  {author} {\bibfnamefont {M.}~\bibnamefont {Dankerl}}, \bibinfo {author}
  {\bibfnamefont {S.}~\bibnamefont {Pezzagna}}, \bibinfo {author}
  {\bibfnamefont {J.}~\bibnamefont {Meijer}}, \bibinfo {author} {\bibfnamefont
  {F.}~\bibnamefont {Jelezko}}, \bibinfo {author} {\bibfnamefont
  {J.}~\bibnamefont {Wrachtrup}}, \bibinfo {author} {\bibfnamefont
  {M.}~\bibnamefont {Stutzmann}}, \bibinfo {author} {\bibfnamefont
  {F.}~\bibnamefont {Reinhard}},\ and\ \bibinfo {author} {\bibfnamefont
  {J.~A.}\ \bibnamefont {Garrido}},\ }\bibfield  {title} {\bibinfo {title}
  {Chemical control of the charge state of nitrogen-vacancy centers in
  diamond},\ }\href {https://doi.org/10.1103/PhysRevB.83.081304} {\bibfield
  {journal} {\bibinfo  {journal} {Phys. Rev. B}\ }\textbf {\bibinfo {volume}
  {83}},\ \bibinfo {pages} {081304(R)} (\bibinfo {year} {2011})}\BibitemShut
  {NoStop}%
\bibitem [{\citenamefont {Ariyaratne}\ \emph {et~al.}(2018)\citenamefont
  {Ariyaratne}, \citenamefont {Bluvstein}, \citenamefont {Myers},\ and\
  \citenamefont {Jayich}}]{Ariyaratne2018}%
  \BibitemOpen
  \bibfield  {author} {\bibinfo {author} {\bibfnamefont {A.}~\bibnamefont
  {Ariyaratne}}, \bibinfo {author} {\bibfnamefont {D.}~\bibnamefont
  {Bluvstein}}, \bibinfo {author} {\bibfnamefont {B.~A.}\ \bibnamefont
  {Myers}},\ and\ \bibinfo {author} {\bibfnamefont {A.~C.~B.}\ \bibnamefont
  {Jayich}},\ }\bibfield  {title} {\bibinfo {title} {Nanoscale electrical
  conductivity imaging using a nitrogen-vacancy center in diamond},\ }\href
  {https://doi.org/10.1038/s41467-018-04798-1} {\bibfield  {journal} {\bibinfo
  {journal} {Nature Communications}\ }\textbf {\bibinfo {volume} {9}},\
  \bibinfo {pages} {2406} (\bibinfo {year} {2018})}\BibitemShut {NoStop}%
\bibitem [{\citenamefont {Lang}\ \emph {et~al.}(2020)\citenamefont {Lang},
  \citenamefont {Häußler}, \citenamefont {Fuhrmann}, \citenamefont
  {Waltrich}, \citenamefont {Laddha}, \citenamefont {Scharpf}, \citenamefont
  {Kubanek}, \citenamefont {Naydenov},\ and\ \citenamefont
  {Jelezko}}]{Lang2020}%
  \BibitemOpen
  \bibfield  {author} {\bibinfo {author} {\bibfnamefont {J.}~\bibnamefont
  {Lang}}, \bibinfo {author} {\bibfnamefont {S.}~\bibnamefont {Häußler}},
  \bibinfo {author} {\bibfnamefont {J.}~\bibnamefont {Fuhrmann}}, \bibinfo
  {author} {\bibfnamefont {R.}~\bibnamefont {Waltrich}}, \bibinfo {author}
  {\bibfnamefont {S.}~\bibnamefont {Laddha}}, \bibinfo {author} {\bibfnamefont
  {J.}~\bibnamefont {Scharpf}}, \bibinfo {author} {\bibfnamefont
  {A.}~\bibnamefont {Kubanek}}, \bibinfo {author} {\bibfnamefont
  {B.}~\bibnamefont {Naydenov}},\ and\ \bibinfo {author} {\bibfnamefont
  {F.}~\bibnamefont {Jelezko}},\ }\bibfield  {title} {\bibinfo {title} {Long
  optical coherence times of shallow-implanted, negatively charged silicon
  vacancy centers in diamond},\ }\href {https://doi.org/10.1063/1.5143014}
  {\bibfield  {journal} {\bibinfo  {journal} {Applied Physics Letters}\
  }\textbf {\bibinfo {volume} {116}},\ \bibinfo {pages} {064001} (\bibinfo
  {year} {2020})}\BibitemShut {NoStop}%
\bibitem [{\citenamefont {Brown}\ \emph {et~al.}(2019)\citenamefont {Brown},
  \citenamefont {Chartier}, \citenamefont {Sweet}, \citenamefont {Hopper},\
  and\ \citenamefont {Bassett}}]{Brown2019}%
  \BibitemOpen
  \bibfield  {author} {\bibinfo {author} {\bibfnamefont {K.~J.}\ \bibnamefont
  {Brown}}, \bibinfo {author} {\bibfnamefont {E.}~\bibnamefont {Chartier}},
  \bibinfo {author} {\bibfnamefont {E.~M.}\ \bibnamefont {Sweet}}, \bibinfo
  {author} {\bibfnamefont {D.~A.}\ \bibnamefont {Hopper}},\ and\ \bibinfo
  {author} {\bibfnamefont {L.~C.}\ \bibnamefont {Bassett}},\ }\bibfield
  {title} {\bibinfo {title} {Cleaning diamond surfaces using boiling acid
  treatment in a standard laboratory chemical hood},\ }\href
  {https://doi.org/10.1016/j.jchas.2019.06.001} {\bibfield  {journal} {\bibinfo
   {journal} {Journal of Chemical Health \& Safety}\ }\textbf {\bibinfo
  {volume} {26}},\ \bibinfo {pages} {40} (\bibinfo {year} {2019})}\BibitemShut
  {NoStop}%
\bibitem [{\citenamefont {Salvadori}\ \emph {et~al.}(2010)\citenamefont
  {Salvadori}, \citenamefont {Araújo}, \citenamefont {Teixeira}, \citenamefont
  {Cattani}, \citenamefont {Pasquarelli}, \citenamefont {Oks},\ and\
  \citenamefont {Brown}}]{Salvadori2010}%
  \BibitemOpen
  \bibfield  {author} {\bibinfo {author} {\bibfnamefont {M.}~\bibnamefont
  {Salvadori}}, \bibinfo {author} {\bibfnamefont {W.}~\bibnamefont {Araújo}},
  \bibinfo {author} {\bibfnamefont {F.}~\bibnamefont {Teixeira}}, \bibinfo
  {author} {\bibfnamefont {M.}~\bibnamefont {Cattani}}, \bibinfo {author}
  {\bibfnamefont {A.}~\bibnamefont {Pasquarelli}}, \bibinfo {author}
  {\bibfnamefont {E.}~\bibnamefont {Oks}},\ and\ \bibinfo {author}
  {\bibfnamefont {I.}~\bibnamefont {Brown}},\ }\bibfield  {title} {\bibinfo
  {title} {Termination of diamond surfaces with hydrogen, oxygen and fluorine
  using a small, simple plasma gun},\ }\href
  {https://doi.org/https://doi.org/10.1016/j.diamond.2010.01.002} {\bibfield
  {journal} {\bibinfo  {journal} {Diamond and Related Materials}\ }\textbf
  {\bibinfo {volume} {19}},\ \bibinfo {pages} {324} (\bibinfo {year}
  {2010})}\BibitemShut {NoStop}%
\bibitem [{\citenamefont {Cui}\ and\ \citenamefont {Hu}(2013)}]{Cui2013}%
  \BibitemOpen
  \bibfield  {author} {\bibinfo {author} {\bibfnamefont {S.}~\bibnamefont
  {Cui}}\ and\ \bibinfo {author} {\bibfnamefont {E.~L.}\ \bibnamefont {Hu}},\
  }\bibfield  {title} {\bibinfo {title} {Increased negatively charged
  nitrogen-vacancy centers in fluorinated diamond},\ }\href
  {https://doi.org/10.1063/1.4817651} {\bibfield  {journal} {\bibinfo
  {journal} {Applied Physics Letters}\ }\textbf {\bibinfo {volume} {103}},\
  \bibinfo {pages} {051603} (\bibinfo {year} {2013})}\BibitemShut {NoStop}%
\bibitem [{\citenamefont {Findler}\ \emph {et~al.}(2020)\citenamefont
  {Findler}, \citenamefont {Lang}, \citenamefont {Osterkamp}, \citenamefont
  {Nesládek},\ and\ \citenamefont {Jelezko}}]{Findler2020}%
  \BibitemOpen
  \bibfield  {author} {\bibinfo {author} {\bibfnamefont {C.}~\bibnamefont
  {Findler}}, \bibinfo {author} {\bibfnamefont {J.}~\bibnamefont {Lang}},
  \bibinfo {author} {\bibfnamefont {C.}~\bibnamefont {Osterkamp}}, \bibinfo
  {author} {\bibfnamefont {M.}~\bibnamefont {Nesládek}},\ and\ \bibinfo
  {author} {\bibfnamefont {F.}~\bibnamefont {Jelezko}},\ }\bibfield  {title}
  {\bibinfo {title} {Indirect overgrowth as a synthesis route for superior
  diamond nano sensors},\ }\href {https://doi.org/10.1038/s41598-020-79943-2}
  {\bibfield  {journal} {\bibinfo  {journal} {Scientific Reports}\ }\textbf
  {\bibinfo {volume} {10}},\ \bibinfo {pages} {22404} (\bibinfo {year}
  {2020})}\BibitemShut {NoStop}%
\bibitem [{\citenamefont {Hacquebard}\ and\ \citenamefont
  {Childress}(2018)}]{Hac2018}%
  \BibitemOpen
  \bibfield  {author} {\bibinfo {author} {\bibfnamefont {L.}~\bibnamefont
  {Hacquebard}}\ and\ \bibinfo {author} {\bibfnamefont {L.}~\bibnamefont
  {Childress}},\ }\bibfield  {title} {\bibinfo {title} {Charge-state dynamics
  during excitation and depletion of the nitrogen-vacancy center in diamond},\
  }\href {https://doi.org/10.1103/PhysRevA.97.063408} {\bibfield  {journal}
  {\bibinfo  {journal} {Phys. Rev. A}\ }\textbf {\bibinfo {volume} {97}},\
  \bibinfo {pages} {063408} (\bibinfo {year} {2018})}\BibitemShut {NoStop}%
\bibitem [{\citenamefont {Waldherr}\ \emph {et~al.}(2011)\citenamefont
  {Waldherr}, \citenamefont {Beck}, \citenamefont {Steiner}, \citenamefont
  {Neumann}, \citenamefont {Gali}, \citenamefont {Frauenheim}, \citenamefont
  {Jelezko},\ and\ \citenamefont {Wrachtrup}}]{Waldherr2011}%
  \BibitemOpen
  \bibfield  {author} {\bibinfo {author} {\bibfnamefont {G.}~\bibnamefont
  {Waldherr}}, \bibinfo {author} {\bibfnamefont {J.}~\bibnamefont {Beck}},
  \bibinfo {author} {\bibfnamefont {M.}~\bibnamefont {Steiner}}, \bibinfo
  {author} {\bibfnamefont {P.}~\bibnamefont {Neumann}}, \bibinfo {author}
  {\bibfnamefont {A.}~\bibnamefont {Gali}}, \bibinfo {author} {\bibfnamefont
  {T.}~\bibnamefont {Frauenheim}}, \bibinfo {author} {\bibfnamefont
  {F.}~\bibnamefont {Jelezko}},\ and\ \bibinfo {author} {\bibfnamefont
  {J.}~\bibnamefont {Wrachtrup}},\ }\bibfield  {title} {\bibinfo {title} {Dark
  states of single nitrogen-vacancy centers in diamond unraveled by single shot
  nmr},\ }\href {https://doi.org/10.1103/PhysRevLett.106.157601} {\bibfield
  {journal} {\bibinfo  {journal} {Phys. Rev. Lett.}\ }\textbf {\bibinfo
  {volume} {106}},\ \bibinfo {pages} {157601} (\bibinfo {year}
  {2011})}\BibitemShut {NoStop}%
\bibitem [{\citenamefont {Kim}\ \emph {et~al.}(2015)\citenamefont {Kim},
  \citenamefont {Mamin}, \citenamefont {Sherwood}, \citenamefont {Ohno},
  \citenamefont {Awschalom},\ and\ \citenamefont {Rugar}}]{Kim2015}%
  \BibitemOpen
  \bibfield  {author} {\bibinfo {author} {\bibfnamefont {M.}~\bibnamefont
  {Kim}}, \bibinfo {author} {\bibfnamefont {H.~J.}\ \bibnamefont {Mamin}},
  \bibinfo {author} {\bibfnamefont {M.~H.}\ \bibnamefont {Sherwood}}, \bibinfo
  {author} {\bibfnamefont {K.}~\bibnamefont {Ohno}}, \bibinfo {author}
  {\bibfnamefont {D.~D.}\ \bibnamefont {Awschalom}},\ and\ \bibinfo {author}
  {\bibfnamefont {D.}~\bibnamefont {Rugar}},\ }\bibfield  {title} {\bibinfo
  {title} {Decoherence of near-surface nitrogen-vacancy centers due to electric
  field noise},\ }\href {https://doi.org/10.1103/PhysRevLett.115.087602}
  {\bibfield  {journal} {\bibinfo  {journal} {Phys. Rev. Lett.}\ }\textbf
  {\bibinfo {volume} {115}},\ \bibinfo {pages} {087602} (\bibinfo {year}
  {2015})}\BibitemShut {NoStop}%
\bibitem [{\citenamefont {Karaveli}\ \emph {et~al.}(2016)\citenamefont
  {Karaveli}, \citenamefont {Gaathon}, \citenamefont {Wolcott}, \citenamefont
  {Sakakibara}, \citenamefont {Shemesh}, \citenamefont {Peterka}, \citenamefont
  {Boyden}, \citenamefont {Owen}, \citenamefont {Yuste},\ and\ \citenamefont
  {Englund}}]{Karaveli2016}%
  \BibitemOpen
  \bibfield  {author} {\bibinfo {author} {\bibfnamefont {S.}~\bibnamefont
  {Karaveli}}, \bibinfo {author} {\bibfnamefont {O.}~\bibnamefont {Gaathon}},
  \bibinfo {author} {\bibfnamefont {A.}~\bibnamefont {Wolcott}}, \bibinfo
  {author} {\bibfnamefont {R.}~\bibnamefont {Sakakibara}}, \bibinfo {author}
  {\bibfnamefont {O.~A.}\ \bibnamefont {Shemesh}}, \bibinfo {author}
  {\bibfnamefont {D.~S.}\ \bibnamefont {Peterka}}, \bibinfo {author}
  {\bibfnamefont {E.~S.}\ \bibnamefont {Boyden}}, \bibinfo {author}
  {\bibfnamefont {J.~S.}\ \bibnamefont {Owen}}, \bibinfo {author}
  {\bibfnamefont {R.}~\bibnamefont {Yuste}},\ and\ \bibinfo {author}
  {\bibfnamefont {D.}~\bibnamefont {Englund}},\ }\bibfield  {title} {\bibinfo
  {title} {Modulation of nitrogen vacancy charge state and fluorescence in
  nanodiamonds using electrochemical potential},\ }\href
  {https://doi.org/10.1073/pnas.1504451113} {\bibfield  {journal} {\bibinfo
  {journal} {Proceedings of the National Academy of Sciences}\ }\textbf
  {\bibinfo {volume} {113}},\ \bibinfo {pages} {3938} (\bibinfo {year}
  {2016})}\BibitemShut {NoStop}%
\bibitem [{\citenamefont {Viola}\ \emph {et~al.}(1999)\citenamefont {Viola},
  \citenamefont {Knill},\ and\ \citenamefont {Lloyd}}]{Viola1999}%
  \BibitemOpen
  \bibfield  {author} {\bibinfo {author} {\bibfnamefont {L.}~\bibnamefont
  {Viola}}, \bibinfo {author} {\bibfnamefont {E.}~\bibnamefont {Knill}},\ and\
  \bibinfo {author} {\bibfnamefont {S.}~\bibnamefont {Lloyd}},\ }\bibfield
  {title} {\bibinfo {title} {Dynamical decoupling of open quantum systems},\
  }\href {https://doi.org/10.1103/PhysRevLett.82.2417} {\bibfield  {journal}
  {\bibinfo  {journal} {Phys. Rev. Lett.}\ }\textbf {\bibinfo {volume} {82}},\
  \bibinfo {pages} {2417} (\bibinfo {year} {1999})}\BibitemShut {NoStop}%
\bibitem [{\citenamefont {Bluvstein}\ \emph
  {et~al.}(2019{\natexlab{b}})\citenamefont {Bluvstein}, \citenamefont {Zhang},
  \citenamefont {McLellan}, \citenamefont {Williams},\ and\ \citenamefont
  {Jayich}}]{Bluvstein2019a}%
  \BibitemOpen
  \bibfield  {author} {\bibinfo {author} {\bibfnamefont {D.}~\bibnamefont
  {Bluvstein}}, \bibinfo {author} {\bibfnamefont {Z.}~\bibnamefont {Zhang}},
  \bibinfo {author} {\bibfnamefont {C.~A.}\ \bibnamefont {McLellan}}, \bibinfo
  {author} {\bibfnamefont {N.~R.}\ \bibnamefont {Williams}},\ and\ \bibinfo
  {author} {\bibfnamefont {A.~C.~B.}\ \bibnamefont {Jayich}},\ }\bibfield
  {title} {\bibinfo {title} {Extending the quantum coherence of a near-surface
  qubit by coherently driving the paramagnetic surface environment},\ }\href
  {https://doi.org/10.1103/PhysRevLett.123.146804} {\bibfield  {journal}
  {\bibinfo  {journal} {Phys Rev Lett}\ }\textbf {\bibinfo {volume} {123}},\
  \bibinfo {pages} {146804} (\bibinfo {year} {2019}{\natexlab{b}})}\BibitemShut
  {NoStop}%
\bibitem [{\citenamefont {Bauch}\ \emph {et~al.}(2018)\citenamefont {Bauch},
  \citenamefont {Hart}, \citenamefont {Schloss}, \citenamefont {Turner},
  \citenamefont {Barry}, \citenamefont {Kehayias}, \citenamefont {Singh},\ and\
  \citenamefont {Walsworth}}]{Bauch2018}%
  \BibitemOpen
  \bibfield  {author} {\bibinfo {author} {\bibfnamefont {E.}~\bibnamefont
  {Bauch}}, \bibinfo {author} {\bibfnamefont {C.~A.}\ \bibnamefont {Hart}},
  \bibinfo {author} {\bibfnamefont {J.~M.}\ \bibnamefont {Schloss}}, \bibinfo
  {author} {\bibfnamefont {M.~J.}\ \bibnamefont {Turner}}, \bibinfo {author}
  {\bibfnamefont {J.~F.}\ \bibnamefont {Barry}}, \bibinfo {author}
  {\bibfnamefont {P.}~\bibnamefont {Kehayias}}, \bibinfo {author}
  {\bibfnamefont {S.}~\bibnamefont {Singh}},\ and\ \bibinfo {author}
  {\bibfnamefont {R.~L.}\ \bibnamefont {Walsworth}},\ }\bibfield  {title}
  {\bibinfo {title} {Ultralong dephasing times in solid-state spin ensembles
  via quantum control},\ }\href {https://doi.org/10.1103/PhysRevX.8.031025}
  {\bibfield  {journal} {\bibinfo  {journal} {Phys. Rev. X}\ }\textbf {\bibinfo
  {volume} {8}},\ \bibinfo {pages} {031025} (\bibinfo {year}
  {2018})}\BibitemShut {NoStop}%
\bibitem [{\citenamefont {Cai}\ \emph {et~al.}(2012)\citenamefont {Cai},
  \citenamefont {Naydenov}, \citenamefont {Pfeiffer}, \citenamefont
  {McGuinness}, \citenamefont {Jahnke}, \citenamefont {Jelezko}, \citenamefont
  {Plenio},\ and\ \citenamefont {Retzker}}]{Cai2012}%
  \BibitemOpen
  \bibfield  {author} {\bibinfo {author} {\bibfnamefont {J.-M.}\ \bibnamefont
  {Cai}}, \bibinfo {author} {\bibfnamefont {B.}~\bibnamefont {Naydenov}},
  \bibinfo {author} {\bibfnamefont {R.}~\bibnamefont {Pfeiffer}}, \bibinfo
  {author} {\bibfnamefont {L.~P.}\ \bibnamefont {McGuinness}}, \bibinfo
  {author} {\bibfnamefont {K.~D.}\ \bibnamefont {Jahnke}}, \bibinfo {author}
  {\bibfnamefont {F.}~\bibnamefont {Jelezko}}, \bibinfo {author} {\bibfnamefont
  {M.~B.}\ \bibnamefont {Plenio}},\ and\ \bibinfo {author} {\bibfnamefont
  {A.}~\bibnamefont {Retzker}},\ }\bibfield  {title} {\bibinfo {title} {Robust
  dynamical decoupling with concatenated continuous driving},\ }\href
  {https://doi.org/10.1088/1367-2630/14/11/113023} {\bibfield  {journal}
  {\bibinfo  {journal} {New Journal of Physics}\ }\textbf {\bibinfo {volume}
  {14}},\ \bibinfo {pages} {113023} (\bibinfo {year} {2012})}\BibitemShut
  {NoStop}%
\bibitem [{\citenamefont {Stark}\ \emph {et~al.}(2017)\citenamefont {Stark},
  \citenamefont {Aharon}, \citenamefont {Unden}, \citenamefont {Louzon},
  \citenamefont {Huck}, \citenamefont {Retzker}, \citenamefont {Andersen},\
  and\ \citenamefont {Jelezko}}]{Stark2017}%
  \BibitemOpen
  \bibfield  {author} {\bibinfo {author} {\bibfnamefont {A.}~\bibnamefont
  {Stark}}, \bibinfo {author} {\bibfnamefont {N.}~\bibnamefont {Aharon}},
  \bibinfo {author} {\bibfnamefont {T.}~\bibnamefont {Unden}}, \bibinfo
  {author} {\bibfnamefont {D.}~\bibnamefont {Louzon}}, \bibinfo {author}
  {\bibfnamefont {A.}~\bibnamefont {Huck}}, \bibinfo {author} {\bibfnamefont
  {A.}~\bibnamefont {Retzker}}, \bibinfo {author} {\bibfnamefont {U.~L.}\
  \bibnamefont {Andersen}},\ and\ \bibinfo {author} {\bibfnamefont
  {F.}~\bibnamefont {Jelezko}},\ }\bibfield  {title} {\bibinfo {title}
  {Narrow-bandwidth sensing of high-frequency fields with continuous dynamical
  decoupling},\ }\href {https://doi.org/10.1038/s41467-017-01159-2} {\bibfield
  {journal} {\bibinfo  {journal} {Nature Communications}\ }\textbf {\bibinfo
  {volume} {8}},\ \bibinfo {pages} {1105} (\bibinfo {year} {2017})}\BibitemShut
  {NoStop}%
\bibitem [{\citenamefont {Stacey}\ \emph {et~al.}(2019)\citenamefont {Stacey},
  \citenamefont {Dontschuk}, \citenamefont {Chou}, \citenamefont {Broadway},
  \citenamefont {Schenk}, \citenamefont {Sear}, \citenamefont {Tetienne},
  \citenamefont {Hoffman}, \citenamefont {Prawer}, \citenamefont {Pakes},
  \citenamefont {Tadich}, \citenamefont {de~Leon}, \citenamefont {Gali},\ and\
  \citenamefont {Hollenberg}}]{Stacey2019}%
  \BibitemOpen
  \bibfield  {author} {\bibinfo {author} {\bibfnamefont {A.}~\bibnamefont
  {Stacey}}, \bibinfo {author} {\bibfnamefont {N.}~\bibnamefont {Dontschuk}},
  \bibinfo {author} {\bibfnamefont {J.-P.}\ \bibnamefont {Chou}}, \bibinfo
  {author} {\bibfnamefont {D.~A.}\ \bibnamefont {Broadway}}, \bibinfo {author}
  {\bibfnamefont {A.~K.}\ \bibnamefont {Schenk}}, \bibinfo {author}
  {\bibfnamefont {M.~J.}\ \bibnamefont {Sear}}, \bibinfo {author}
  {\bibfnamefont {J.-P.}\ \bibnamefont {Tetienne}}, \bibinfo {author}
  {\bibfnamefont {A.}~\bibnamefont {Hoffman}}, \bibinfo {author} {\bibfnamefont
  {S.}~\bibnamefont {Prawer}}, \bibinfo {author} {\bibfnamefont {C.~I.}\
  \bibnamefont {Pakes}}, \bibinfo {author} {\bibfnamefont {A.}~\bibnamefont
  {Tadich}}, \bibinfo {author} {\bibfnamefont {N.~P.}\ \bibnamefont {de~Leon}},
  \bibinfo {author} {\bibfnamefont {A.}~\bibnamefont {Gali}},\ and\ \bibinfo
  {author} {\bibfnamefont {L.~C.~L.}\ \bibnamefont {Hollenberg}},\ }\bibfield
  {title} {\bibinfo {title} {Evidence for primal sp$^{2}$ defects at the
  diamond surface: Candidates for electron trapping and noise sources},\ }\href
  {https://doi.org/https://doi.org/10.1002/admi.201801449} {\bibfield
  {journal} {\bibinfo  {journal} {Advanced Materials Interfaces}\ }\textbf
  {\bibinfo {volume} {6}},\ \bibinfo {pages} {1801449} (\bibinfo {year}
  {2019})}\BibitemShut {NoStop}%
\bibitem [{\citenamefont {Giri}\ \emph {et~al.}(2018)\citenamefont {Giri},
  \citenamefont {Gorrini}, \citenamefont {Dorigoni}, \citenamefont {Avalos},
  \citenamefont {Cazzanelli}, \citenamefont {Tambalo},\ and\ \citenamefont
  {Bifone}}]{Giri2018}%
  \BibitemOpen
  \bibfield  {author} {\bibinfo {author} {\bibfnamefont {R.}~\bibnamefont
  {Giri}}, \bibinfo {author} {\bibfnamefont {F.}~\bibnamefont {Gorrini}},
  \bibinfo {author} {\bibfnamefont {C.}~\bibnamefont {Dorigoni}}, \bibinfo
  {author} {\bibfnamefont {C.~E.}\ \bibnamefont {Avalos}}, \bibinfo {author}
  {\bibfnamefont {M.}~\bibnamefont {Cazzanelli}}, \bibinfo {author}
  {\bibfnamefont {S.}~\bibnamefont {Tambalo}},\ and\ \bibinfo {author}
  {\bibfnamefont {A.}~\bibnamefont {Bifone}},\ }\bibfield  {title} {\bibinfo
  {title} {Coupled charge and spin dynamics in high-density ensembles of
  nitrogen-vacancy centers in diamond},\ }\href
  {https://doi.org/10.1103/PhysRevB.98.045401} {\bibfield  {journal} {\bibinfo
  {journal} {Phys. Rev. B}\ }\textbf {\bibinfo {volume} {98}},\ \bibinfo
  {pages} {045401} (\bibinfo {year} {2018})}\BibitemShut {NoStop}%
\bibitem [{\citenamefont {Sangtawesin}\ \emph {et~al.}(2019)\citenamefont
  {Sangtawesin}, \citenamefont {Dwyer}, \citenamefont {Srinivasan},
  \citenamefont {Allred}, \citenamefont {Rodgers}, \citenamefont {De~Greve},
  \citenamefont {Stacey}, \citenamefont {Dontschuk}, \citenamefont {O'Donnell},
  \citenamefont {Hu}, \citenamefont {Evans}, \citenamefont {Jaye},
  \citenamefont {Fischer}, \citenamefont {Markham}, \citenamefont {Twitchen},
  \citenamefont {Park}, \citenamefont {Lukin},\ and\ \citenamefont
  {de~Leon}}]{Sangtawesin2019}%
  \BibitemOpen
  \bibfield  {author} {\bibinfo {author} {\bibfnamefont {S.}~\bibnamefont
  {Sangtawesin}}, \bibinfo {author} {\bibfnamefont {B.~L.}\ \bibnamefont
  {Dwyer}}, \bibinfo {author} {\bibfnamefont {S.}~\bibnamefont {Srinivasan}},
  \bibinfo {author} {\bibfnamefont {J.~J.}\ \bibnamefont {Allred}}, \bibinfo
  {author} {\bibfnamefont {L.~V.~H.}\ \bibnamefont {Rodgers}}, \bibinfo
  {author} {\bibfnamefont {K.}~\bibnamefont {De~Greve}}, \bibinfo {author}
  {\bibfnamefont {A.}~\bibnamefont {Stacey}}, \bibinfo {author} {\bibfnamefont
  {N.}~\bibnamefont {Dontschuk}}, \bibinfo {author} {\bibfnamefont {K.~M.}\
  \bibnamefont {O'Donnell}}, \bibinfo {author} {\bibfnamefont {D.}~\bibnamefont
  {Hu}}, \bibinfo {author} {\bibfnamefont {D.~A.}\ \bibnamefont {Evans}},
  \bibinfo {author} {\bibfnamefont {C.}~\bibnamefont {Jaye}}, \bibinfo {author}
  {\bibfnamefont {D.~A.}\ \bibnamefont {Fischer}}, \bibinfo {author}
  {\bibfnamefont {M.~L.}\ \bibnamefont {Markham}}, \bibinfo {author}
  {\bibfnamefont {D.~J.}\ \bibnamefont {Twitchen}}, \bibinfo {author}
  {\bibfnamefont {H.}~\bibnamefont {Park}}, \bibinfo {author} {\bibfnamefont
  {M.~D.}\ \bibnamefont {Lukin}},\ and\ \bibinfo {author} {\bibfnamefont
  {N.~P.}\ \bibnamefont {de~Leon}},\ }\bibfield  {title} {\bibinfo {title}
  {Origins of diamond surface noise probed by correlating single-spin
  measurements with surface spectroscopy},\ }\href
  {https://doi.org/10.1103/PhysRevX.9.031052} {\bibfield  {journal} {\bibinfo
  {journal} {Phys. Rev. X}\ }\textbf {\bibinfo {volume} {9}},\ \bibinfo {pages}
  {031052} (\bibinfo {year} {2019})}\BibitemShut {NoStop}%
\bibitem [{\citenamefont {Grotz}\ \emph {et~al.}(2012)\citenamefont {Grotz},
  \citenamefont {Hauf}, \citenamefont {Dankerl}, \citenamefont {Naydenov},
  \citenamefont {Pezzagna}, \citenamefont {Meijer}, \citenamefont {Jelezko},
  \citenamefont {Wrachtrup}, \citenamefont {Stutzmann}, \citenamefont
  {Reinhard},\ and\ \citenamefont {Garrido}}]{Grotz2012}%
  \BibitemOpen
  \bibfield  {author} {\bibinfo {author} {\bibfnamefont {B.}~\bibnamefont
  {Grotz}}, \bibinfo {author} {\bibfnamefont {M.~V.}\ \bibnamefont {Hauf}},
  \bibinfo {author} {\bibfnamefont {M.}~\bibnamefont {Dankerl}}, \bibinfo
  {author} {\bibfnamefont {B.}~\bibnamefont {Naydenov}}, \bibinfo {author}
  {\bibfnamefont {S.}~\bibnamefont {Pezzagna}}, \bibinfo {author}
  {\bibfnamefont {J.}~\bibnamefont {Meijer}}, \bibinfo {author} {\bibfnamefont
  {F.}~\bibnamefont {Jelezko}}, \bibinfo {author} {\bibfnamefont
  {J.}~\bibnamefont {Wrachtrup}}, \bibinfo {author} {\bibfnamefont
  {M.}~\bibnamefont {Stutzmann}}, \bibinfo {author} {\bibfnamefont
  {F.}~\bibnamefont {Reinhard}},\ and\ \bibinfo {author} {\bibfnamefont
  {J.~A.}\ \bibnamefont {Garrido}},\ }\bibfield  {title} {\bibinfo {title}
  {Charge state manipulation of qubits in diamond},\ }\href
  {https://doi.org/10.1038/ncomms1729} {\bibfield  {journal} {\bibinfo
  {journal} {Nature Communications}\ }\textbf {\bibinfo {volume} {3}},\
  \bibinfo {pages} {729} (\bibinfo {year} {2012})}\BibitemShut {NoStop}%
\end{thebibliography}%
\end{document}